\documentclass[aps,prb,twocolumn,floatfix,footinbib,showpacs,superscriptaddress]{revtex4-1}

\usepackage{times}
\usepackage{color}
\usepackage{graphicx}
\usepackage{dcolumn}
\usepackage{amssymb}
\usepackage{amsmath}
\usepackage{amsfonts}
\usepackage{docs}
\usepackage{bm}
\usepackage{tikz}
\usetikzlibrary{shapes}
\usepackage[colorlinks=true,linkcolor=blue,pagecolor=blue,
filecolor=blue,menucolor=blue,urlcolor=blue,citecolor=blue,
anchorcolor=blue]{hyperref}

\newcommand*{\rttensor}[1]{\overline{\overline{#1}}}
\newcommand{\var}[1]{{\operatorname{\mathit{#1}}}}
 \DeclareMathOperator\arctanh{arctanh}
\DeclareRobustCommand*\circled[1]{\tikz[baseline=(char.base)]{
\node[shape=circle,draw,inner sep=0.75pt] (char) {#1};}}
\begin{document}

\title{Epsilon-Near-Zero Response and Tunable Perfect Absorption in Weyl Semimetals}
\author{Klaus Halterman} 
\affiliation{Michelson Lab, Physics Division, Naval Air Warfare Center, China Lake, California 93555}
\author{Mohammad Alidoust} 
\affiliation{Department of Physics, K.N. Toosi University of Technology, Tehran 15875-4416, Iran}
\author{Alexander Zyuzin}
   \affiliation{Department of Applied Physics, Aalto University, P. O. Box 15100, FI-00076 AALTO, Finland}
  \affiliation{ Ioffe Physical-Technical Institute, St. Petersburg 194021, Russia}

\begin{abstract}
We theoretically study the electromagnetic response of type-I and type-II
centrosymmetric Weyl metals. We derive an anisotropic permittivity tensor with off-diagonal elements to 
describe such gyrotropic media. 
Our findings reveal that for appropriate Weyl cones tilts, the real part of the 
transverse component of the permittivity can exhibit an epsilon-near-zero response. 
The tilt parameter can
also control the amount of loss in the medium, ranging from lossless
to dissipative when transitioning from type-I to
type-II. 
Similarly, by tuning either the frequency 
 of the electromagnetic field or the chemical potential in the system, 
an epsilon-near-zero response can appear as the permittivity of the
Weyl semimetal
transitions between positive and negative values. 
Employing the obtained permittivity tensor, we consider a setup where the Weyl semimetal is deposited on a perfect conductive substrate and study the refection and absorption characteristics of this layered configuration. 
We show that by choosing the proper geometrical and material parameters, devices
can be created
that perfectly absorb electromagnetic energy over a wide angular range of 
incident electromagnetic waves.   
\end{abstract}

\date{\today}

\maketitle

\section{Introduction}\label{intro}

Since the advent of metamaterials, 
efforts to design materials that can be used to 
control  
electromagnetic (EM) fields has flourished.
One prominent route involves developing anisotropic structures whose
EM response  is described by a permittivity  tensor $\rttensor{\epsilon}$ that has components with
extreme values, including epsilon-near-zero (ENZ)\cite{enghetta}
media, where  the real part of a component of $\rttensor{\epsilon}$  vanishes
along a given coordinate axis.
Within the ENZ regime, the phase of the entering 
EM wave can be uniform, and the wavefront
conforms to the shape of the exit side of the ENZ medium.\cite{alu}
 A number of ENZ-based architectures have been  fabricated, including
sub-wavelength dielectric coatings that control  the resonant coupling of light
 with ENZ regions. \cite{colors}
Experimental work with  microwave waveguides\cite{alu2,liu} demonstrated how
 a narrow ENZ channel 
 can lead to 
enhanced electromagnetic coupling.
 A  nanoparticle mixture containing  dielectric and metal constituents 
with an effective ENZ response 
exhibited an  increase in  the superconducting 
critical temperature. \cite{smol}
The propagation of a transverse magnetic optical beam through 
a subwavelength slit demonstrated a
transmission enhancement\cite{funnel} when the
InAsSb semiconductor substrate
was tuned to its ENZ 
frequency.

Many pathways have been studied that lead to the creation of ENZ materials, including intricate combinations
of  metal-dielectric multilayers and arrays of rods, or transparent conducting oxides.
Other approaches involve the use of more exotic materials like graphene \cite{matt}
with its intrinsic two dimensionality and linear dispersion around the Dirac point.
Recently, Weyl semimetals\cite{fang1,burkov,wyl5,wyl6,weyl,rev1,rev2,burkov,wyl3}
 have been added to the
ever expanding class of materials that have useful EM properties.
 The band structure of a Weyl semimetal (WS) 
 is characterized by a conical energy spectrum with  an even number of
 Weyl nodes 
 that are
topologically protected.
The chiralities of Weyl
nodes correspond to  topological charges 
that result in
monopoles and anti-monopoles in the Berry curvature
\cite{wyl5,wyl6}. Indeed, the Weyl semimetal phase manifests
itself  in unusual
surface states with Fermi arcs and chiral anomalies
\cite{wyl5,wyl6,weyl,burkov,wyl3}. 
Weyl semimetals are topologically nontrivial materials that are predicted for the magnetic compounds Y$_2$IrO$_7$, Eu$_2$IrO$_7$, and
HgCr$_2$Se$_4$ \cite{Savrasov,fang2}, and in some
nonmagnetic samples,
including TaAs, TaP, NbAs, and NbP \cite{hsn3,dai,yan,hsn4,qiu1}.
The WS TaAs was shown to have a wide spectral range 
as a room temperature photodetector. \cite{china}

The synthesis of different alloys into 
Weyl semimetal crystals can
result in
a novel type of Weyl semimetal 
that is characterized by titled Weyl nodes and an
open Fermi surface.
This class of Weyl semimetals is 
identified as type-II if the tilt of the Weyl cone
exceeds the Fermi velocity.
\cite{typ2_1,typ2_2,typ2_3,typ2_4,typ2_5,typ2_6,typ2_8,typ2_7}
Since condensed matter systems do
not require Lorentz invariance,
Weyl semimetals are not restricted to
 closed point-like Fermi surfaces,  and 
 support type-II Weyl fermions
\cite{typ2_6,typ2_1}. 
This new type of Weyl fermion appears at the
boundary between electron and hole pockets \cite{typ2_6,typ2_1}.
The
experimental signatures of this new phase were 
recently reported in Refs.~$\,$\onlinecite{typ2_3,typ2_4}. The  effects of Weyl cone tilt on the  optical conductivity and polarization 
was recently studied.\cite{collect} 
The effect of a tilt on the absorption
of circular polarized light was studied for both 
 type-I and type-II cases and it was shown that reversing the tilting direction of Weyl nodes the right-hand and left-hand responses of Weyl semimetal becomes reversed.\cite{muk1} Also, It was found that chirality or the tilt-sign in Weyl semimetals with tilted cones in the absence of time-reversal and inversion symmetries 
 can change the sign of the Weyl contribution to the absorptive Hall conductivity \cite{muk2}.

Tunable metamaterial absorbers with active materials have
been explored with a variety of different materials.\cite{kats,chen,shrek,shank}
The broad tunability of the chemical potential in a WS 
makes
it a promising material for
photonics and plasmonics applications\cite{qi1,zhou1,Kharzeev1,Hofmann1,zyuzin,wang,Sushkov1,chin,hau,tim,valla,wang,chin,hau,tim}. 
The chiral anomaly in a WS
can alter surface plasmons and
the EM response \cite{qi1,zhou1,Kharzeev1,Hofmann1,zyuzin,wang,Sushkov1}.
It has been shown theoretically \cite{Sushkov1}  that
 measurements of the optical conductivity 
and the  temperature dependence of the 
free carrier response in pyrochlore
Eu$_2$Ir$_2$O$_7$ is consistent with the WS phase. 
Also, the interband optical conductivity
reduces to zero in a continuous fashion
at low frequencies as predicted for a WS. 
The
analysis of
experimental data resulted in a Fermi velocity on the order of 
$v_F \approx 4
\times 10^7$ cm/s \cite{wang,Sushkov1}. The surface
magnetoplasmons
of a Weyl semimetal can turn to  low-loss localized guided
modes when two crystals  of the WSs
with different magnetization orientations are
connected \cite{zyuzin}.

In this paper, we study the anisotropic  electromagnetic response of both type-I and type-II Weyl semimetals. 
 Our study includes both analytic and numerical results that reveal the behavior 
 of each component of the dielectric tensor as a function of
the  Weyl cone tilt, chemical potential, and EM wave frequency.
We show that by appropriately tailoring these system parameters, the real part of the 
transverse component of the permittivity can achieve an ENZ response. 
In parallel, we  also demonstrate how the dissipative effects in the medium can be controlled. 
Utilizing the derived permittivity tensor and its subsequent numerical analysis, 
we consider a Weyl semimetal (both type-I and type-II) on top of a perfectly conducting substrate, 
and study the  absorptance of an incident electromagnetic wave from 
vacuum into the Weyl semimetal surface. Solving  
Maxwell's equations, we derive the
reflection and absorption coefficients, and show that by properly choosing 
material and geometric parameters, tunable coherent perfect absorption is feasible over a wide range of incident angles.

The paper is organized as follows. In Sec. \ref{sec1:method}, 
we present the derived permittivity tensor applicable to both type-I and type-II WSs. We apply various approximations, and 
discuss the EM response of
type-I and type-II WSs in Subsec.~\ref{subsec1:zerotilt} and \ref{subsec1:fintlit}, respectively. 
In Subsec.~\ref{subsec1:epsilon}, we numerically illustrate and analyze various features of both types of WSs. 
In Sec. \ref{sec2:abs_rfl}, we present a practical application of the analyses given in Sec.~\ref{sec1:method}. 
Particularly, we study the electromagnetic response of a WS grounded by a perfect conductor. 
Starting from Maxwell's equations, we derive analytical expressions for the reflection and absorption 
coefficients of this structure
as a function of incident electromagnetic wave angle and  thickness of the WS. 
Furthermore, we numerically analyze various aspects of the absorption characteristics of this system. 
Finally, we give concluding remarks in Sec.~\ref{conclusion}. 

\section{ approach and results}\label{sec1:method} 
In this section, we outline the model  Hamiltonian, and 
calculate  
 the 
permittivity tensor  for both a type-I and type-II WS.
General expressions are given for determining each of the 
permittivity components, and
analytic results are derived for various
 limiting cases.
Results are then presented for the susceptibility and epsilon-near-zero
response as a  function of frequency $\omega$, chemical potential
$\mu$, and  tilt parameter $\beta$.
 
\subsection{Permittivity tensor}\label{subsec1:permiti}
Throughout this paper we focus on a model Hamiltonian with broken time
reversal symmetry and only two Weyl nodes. 
This model can be achieved through the stacking of multiple thin films involving a 
topological insulator and 
 ferromagnet blocks, as first proposed theoretically\cite{burkov}. 
The Hamiltonian describing the low energy physics around the two Weyl nodes,
defined by ``$s=\pm$'', is given by: 
\begin{equation}
H_{s}(\mathbf{p}) = v_F[\beta_s (p_z-sQ) +s {\bm \sigma}(\textbf{p}-sQ\textbf{e}_z)].
\end{equation}
Here  $\textbf{e}_z$ is
the unit vector along the $z$ direction, 
and we take the Fermi velocity $v_F$ to be positive. 
The separation between two Weyl points in the $z$ direction in momentum space is defined by $2|Q|$,
where the sign of $Q$ depends on the sign of the magnetization. 
The quantities  $\beta_{\pm}$ are tilting parameters that control the
transition between the type-I and type-II phases. 
For centrosymmetric materials with broken time reversal symmetry, 
we apply the condition  $\beta_+ = - \beta_-$. 
The corresponding electron Green functions are given by
\begin{equation}
G_s(\varepsilon_n,\textbf{p})=\frac{1}{2}\sum\limits_{t=\pm 1}
\frac{1+st \boldsymbol{\sigma}\mathbf{p}^{(s)}/|\mathbf{p}^{(s)}| }
{i\varepsilon_n+\mu-v_F[\beta_sp^{(s)}_{z}+t|\textbf{p}^{(s)}|]},~~~
\end{equation}
where the index $t$ 
identifies each of the two subbands, 
$\mathbf{p}^{(s)} = \mathbf{p}-sQ\mathbf{e}_z$, and
$\mu$ is the chemical potential, in which we set $\mu\geq 0$ without loss of generality. The fermionic
Matsubara frequency is $\varepsilon_n=\pi
T(2n+1)$, in which $T$ represents temperature and
$n$ is an integer.
The dielectric function $\epsilon_{ab}$($\omega$) (where
$a,b \equiv x,y,z$) is defined through the optical conductivity $\sigma_{ab}$ via:
\begin{align}
\sigma_{ab}(\omega)&=\frac{i}{\omega}\underset{|\textbf{q}|\rightarrow
  0}{\text{lim}}[\Pi_{ab}(\omega,\textbf{q})-
  \Pi_{ab}(0,\textbf{q})],\nonumber \\  
	\epsilon_{ab}(\omega) &=\delta_{ab}+i\frac{\sigma_{ab}(\omega)}{{\epsilon_0}\omega},
\end{align}
where $\epsilon_0$ is the permittivity of free space and the current-current correlation function reads:
\begin{eqnarray}
&&
  \Pi_{ab}(\omega,\textbf{q})=
e^2T\sum\limits_n\sum\limits_{s=\pm}\text{Tr}\int\frac{d^3p}{(2\pi)^3}J_{a,s}G_s(\varepsilon_n+\omega_k,\textbf{p}+\textbf{q})\nonumber\\&&\times J_{b,s}
G_{s}(\varepsilon_n,\textbf{p})\Big\}\Big|_{i\omega_k\rightarrow\omega+i\delta}, \end{eqnarray}
where $\mathbf{J}_{s} = v_F(\beta_s\mathbf{e}_z+s\boldsymbol{\sigma})$, $\omega_k=2\pi Tk$ is the bosonic Matsubara frequency, and $k$ is an integer.
Thus, the permittivity tensor $\rttensor{\epsilon}$ takes the
following gyrotropic form that is valid for both type-I and type-II Weyl semimetal
phases: 
\begin{subequations}
\begin{eqnarray}
\rttensor{\epsilon}
=\left( \begin{array}{ccc}
\epsilon_{xx}(\omega) & \epsilon_{xy}(\omega) & 0 \\
\epsilon_{yx}(\omega) & \epsilon_{yy}(\omega)  & 0 \\
0 &0 &\epsilon_{zz}(\omega)  \end{array}\right),
\end{eqnarray}
where  the off-diagonal components 
are given by $\epsilon_{xy}(\omega)=-\epsilon_{yx}(\omega)=i\gamma (\omega)$. 
These terms
can lead to modified polarization rotations via the
Kerr and Faraday effects.\cite{karg,Kotov1}
Variations in the gyrotropic term can also cause shifts in
the surface plasmon frequency.\cite{berg}
The $\epsilon_{xx,yy}(\omega)$ components are equal and can
be written analytically as, 
\begin{equation}
\label{dispexx}
\epsilon_{xx,yy}(\omega)=1+
\frac{\alpha}{3\pi}
\Biggl[\text{ln} \left|\dfrac{4\Gamma^2}{4 \mu^2-\omega^2}\right| -\dfrac{4 \mu^2}{\omega^2}+i\pi\Theta(\omega-2 \mu)\Biggr],
\end{equation}
in which $\alpha=e^2/(4\pi\epsilon_0 \hbar v_F)$ and  $\Gamma\sim v_F|Q|$, such that $\Gamma\gg (\omega, \mu)$ is the high-energy cut-off where applicability of the linear model no longer holds. Here $\Theta(X)$ represents the usual step function.
The remaining components involve integrals, which  for $\Gamma\gg (\omega, \mu)$, are written in the limit of zero temperature:
\begin{widetext}
\begin{equation}
\label{dispezz}
\epsilon_{zz}(\omega)=1-\frac{\pi\alpha}{\omega^2}\sum\limits_{t=\pm
  1}\sum\limits_{s=\pm}\int\frac{d^3p}{(2\pi)^3}
  \left\{-\frac{p_{\perp}^2t\omega_k^2/4 }{p^3(p^2+\omega_k^2/4)} \Theta(\mu-\zeta_{s,t})+ \left(\beta_s+t\frac{p_z}{p}\right)^2\delta(\mu-\zeta_{s,t}) \right\}\Big|_{i\omega_k\rightarrow\omega+i\delta}.
\end{equation}
It is convenient to separate Fermi surface and vacuum contributions to $\gamma(\omega)=\sum_{s=\pm}[\gamma^{(s)}_{FS}(\omega)+\gamma^{(s)}_0(\omega)]$, where

\begin{eqnarray}\label{dispgam}
\gamma^{(s)}_{FS} (\omega)&=&
\frac{s \alpha}{\omega^2}\int_{0}^{\infty}\frac{p_{\perp}dp_{\perp}}{2\pi} \int_{-\Gamma-sv_FQ}^{\Gamma-sv_FQ}dp_z\frac{p_z}{p}\frac{i\omega_k}{p^2+\omega_k^2/4}[\Theta(\mu-\zeta_{s,+})-\Theta(\mu-\zeta_{s,-})-1]\Big|_{i\omega_k\rightarrow\omega+i\delta},\\
\gamma^{(s)}_{0} (\omega)&=&\frac{s \alpha}{\omega^2}\int_{0}^{\infty}\frac{p_{\perp}dp_{\perp}}{2\pi}\int_{-\Gamma_{0}-sv_FQ}^{\Gamma_{0}-sv_FQ}dp_z\frac{p_z}{p}\frac{i\omega_k}{p^2+\omega_k^2/4}\Big|_{i\omega_k\rightarrow\omega+i\delta},
\end{eqnarray}
\end{widetext} 
\end{subequations}
in which we have defined $\zeta_{s,t}\equiv tp+p_z\beta_s$,  
$p=\sqrt{p_z^2+p_{\perp}^2}$, and a momentum
cutoff along the $z$ axis, $\Gamma$. Generally, the cut-off $\Gamma$ is a function of the tilt parameter. 
Nevertheless, in our calculations, we choose a large enough cut-off and neglect the contribution of $\beta$ to $\Gamma$. 
The cut-off $\Gamma_0>v_F|Q|$ is introduced for the
correct definition of the vacuum contribution.

\subsection{ Zero tilt phase: {\normalsize $\beta=0$}}\label{subsec1:zerotilt}
It is evident that $\epsilon_{xx,yy}(\omega)$ are independent of the tilting
parameters $\beta_{s}$. To reduce $\rttensor{\epsilon}$ to
 the situation
where the Weyl nodes experience no tilt, it suffices to set
$|\beta_{+}|=|\beta_{-}|=0$. 
In this case, $\epsilon_{zz}(\omega)=\epsilon_{xx,yy}(\omega)$,
and the off-diagonal frequency dependent 
component $\gamma(\omega)$ reduces to,
\begin{eqnarray} \label{gamxy}
\gamma(\omega) =\frac{2\alpha}{\pi} \frac{v_F Q}{\omega},
\end{eqnarray}
The gyrotropic parameter $\gamma(\omega)$ 
can play an important role in changing the polarization state of
 electromagnetic waves interacting
 with
  the WS via Faraday and Kerr rotations.\cite{karg} The imaginary term in Eq.~(\ref{dispexx})
describes the interband contribution to the optical
conductivity, which exists only when
 the frequency $\omega$ of the EM wave
satisfies  $ \omega>2\mu$. 
The interval of frequencies in which 
$\epsilon_{xx,yy}$ is real and positive correspond to,
\begin{align}
2\mu> \omega>2\mu\sqrt{\frac{\alpha}{3\pi(1+\frac{2\alpha}{3\pi}\text{ln}\big|\frac{\Gamma}{\mu}\big|)}}.
\end{align}
Thus, for frequencies around the chemical potential, $2\mu> \omega\gtrsim\mu$,
the diagonal components $\epsilon_{xx,yy}$
do not contribute to  dissipation in the medium.
On the other hand, if
$ \omega>2\mu$,  
$\epsilon_{xx,yy}$ acquires an imaginary part,
leading to dissipation.

\subsection{Finite tilt phase: {\normalsize $\beta\neq 0$}}\label{subsec1:fintlit}
We now examine some limiting cases for the integrals in Eqs.~(\ref{dispezz}) and (\ref{dispgam}) when the tilting parameter is nonzero.        
At the charge neutrality point, $\mu=0$, the off-diagonal gyrotropic component $\gamma(\omega)$ reduces to
\begin{equation} \label{gamef0}
\gamma(\omega)=
\frac{\alpha}{\pi}\frac{v_FQ}{\omega}\Big[ \text{min}(1,|\beta_+|^{-1})+\text{min}(1,|\beta_-|^{-1}) \Big].
\end{equation}
It is seen that $\gamma(\omega)$ is independent of the tilt parameter if $|\beta_{s}| < 1$, 
while inversely proportional to it when $|\beta_{s}| > 1$. 
The $\epsilon_{zz}$ component of the dielectric tensor in the limit of zero chemical potential, $\mu=0$, 
can be calculated exactly as well. For $|\beta_{s}| < 1$, we find,
\begin{align}
\label{ezz_ef0}
\epsilon_{zz}=1+\frac{\alpha}{3\pi}\Big[\ln \left|\frac{4\Gamma^2}{\omega^2}\right|+i\pi\Theta(\omega)\Big], 
\end{align}
and when the tilt parameter satisfies $|\beta_{s}|>1$, we arrive at
\begin{align}
\label{ezz2}
\epsilon_{zz}&=1+\frac{\alpha}{3\pi}\Big[\ln \bigg|\frac{4\Gamma^2}{\omega^2}\bigg|+i\pi\Theta(\omega)\Big]\sum\limits_{s=\pm}\frac{1}{4|\beta_s|}\bigg(3-\frac{1}{\beta_s^2}\bigg)\nonumber \\
&-\frac{\alpha \Gamma^2}{\pi\omega^2}\sum\limits_{s=\pm} \big|\beta_s\big| \Big(1-\frac{1}{\beta_s^2} \Big)^2.
\end{align}
Here the last term describes the contribution from the Fermi pocket, which is bounded by $\Gamma/v$ along the $z$ axis in momentum space.
Thus demonstrating that the real part of $\epsilon_{zz}$ is always positive, and independent of the tilt
parameters if $|\beta_{s}| < 1$. Only when the conical
tilt parameters exceed unity ($|\beta_{s}| > 1$) can
the real part of $\epsilon_{zz}$ become 
negative or zero, thus allowing for the possibility of an ENZ response. 
Note that  the amplitude of the imaginary component declines monotonically with increasing the tilt parameter $\beta$.

\begin{figure*}[!tbp]
\centering
\includegraphics[clip, trim=0.2cm 0.2cm 3.5cm 0.2cm, width=0.33\textwidth]{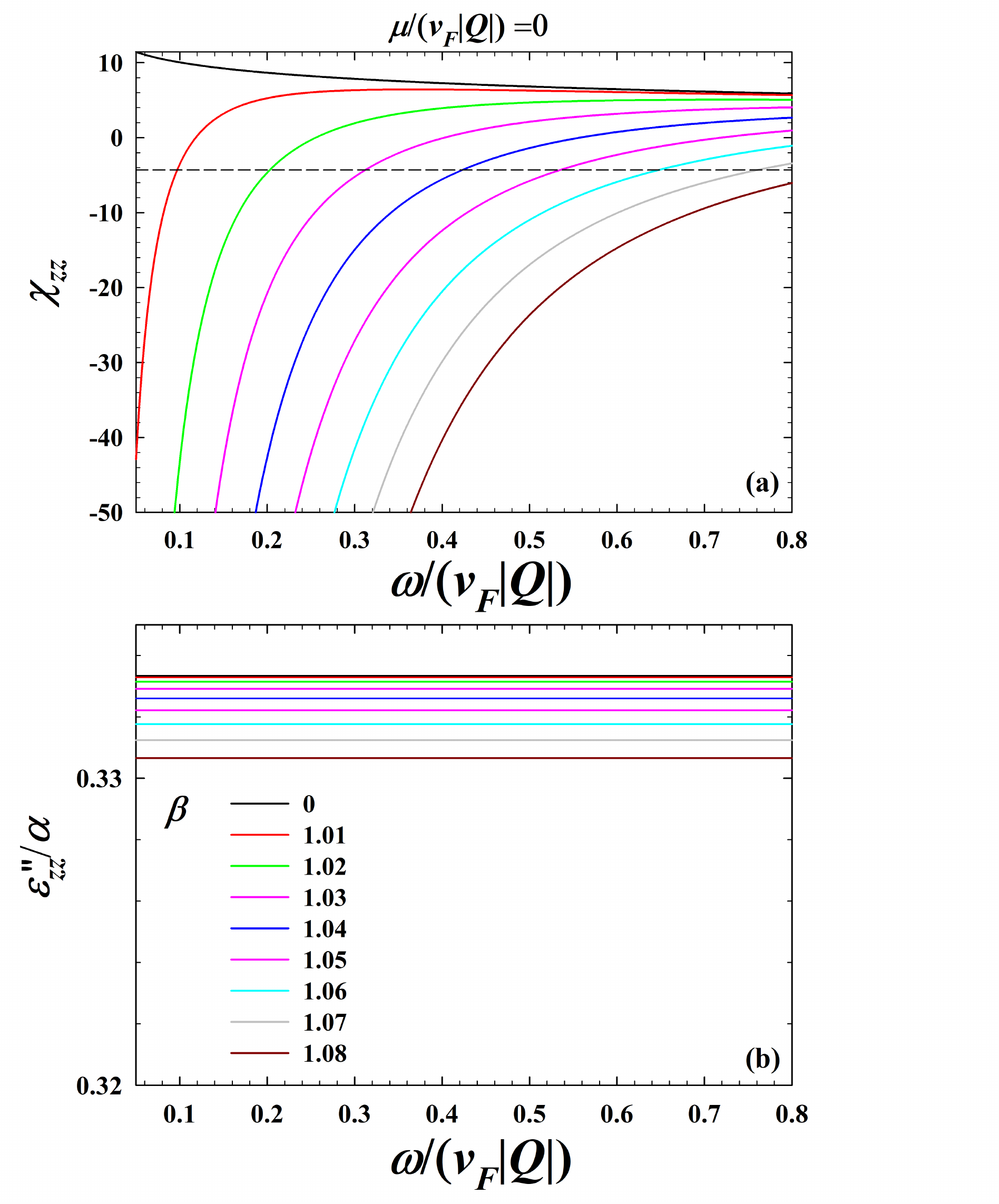}\hspace{-.05cm}
\includegraphics[clip, trim=0.2cm 0.2cm 3.5cm 0.2cm, width=0.33\textwidth]{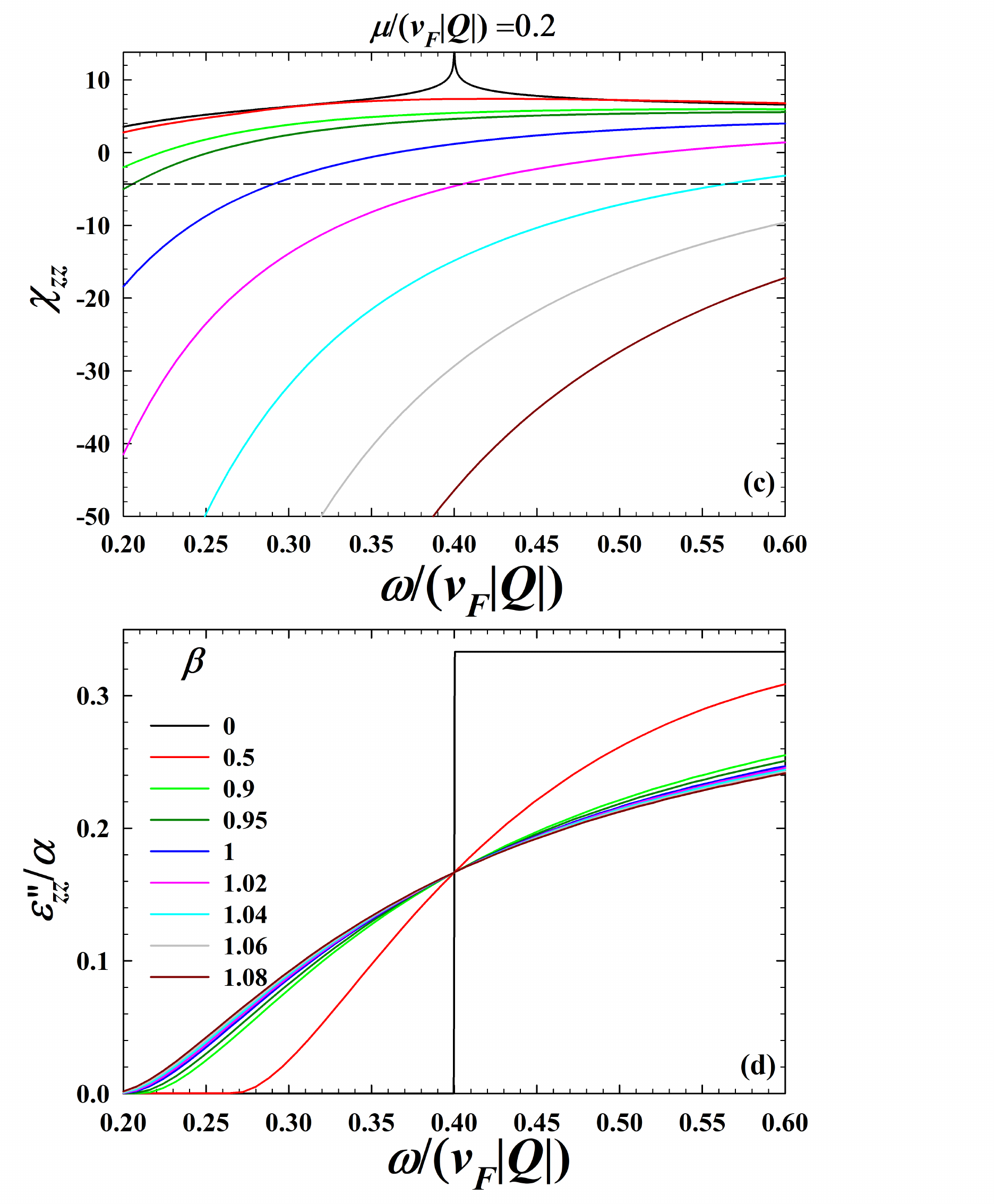}\hspace{-.19cm}
\includegraphics[clip, trim=0.2cm 0.2cm 3.5cm 0.2cm, width=0.33\textwidth]{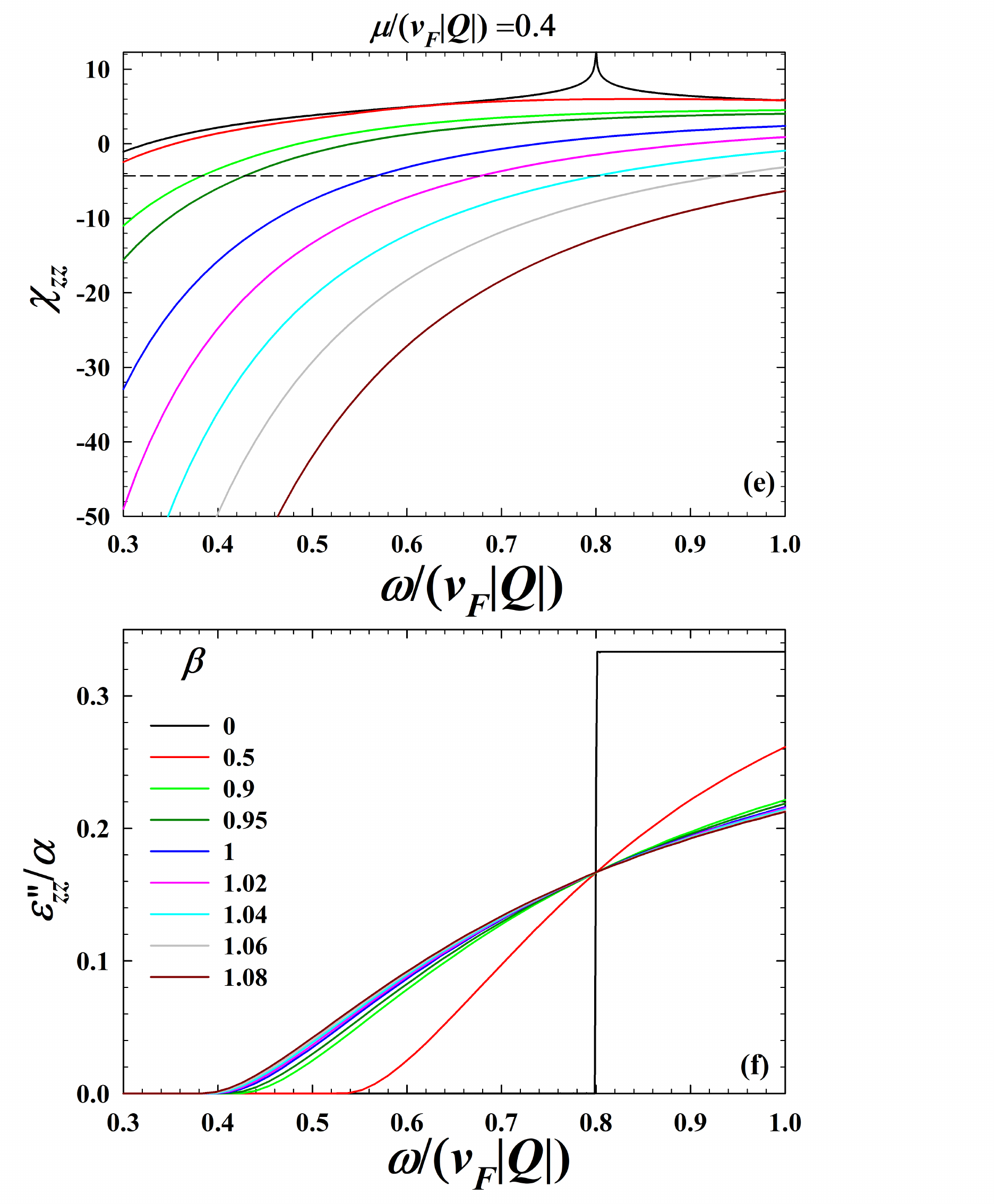}
\caption{(Color online). 
Representative  
behavior of  the permittivity ${\epsilon}_{zz}$
as a function of  frequency. 
The top row corresponds to the normalized 
susceptibility  $\chi_{zz}$, 
while the bottom row is the imaginary component $\epsilon_{zz}''/\alpha$.
The dashed line identifies points where an ENZ response arises.
The tilting of the Weyl cones are seen to have a significant effect 
in creating an ENZ response. The corresponding $\beta$ are
depicted in the legends.
The effects of altering the chemical potential is also shown, where
each column depicts one of the three different $\mu$ considered. In (a)-(b) $\mu/(v_F|Q|)=0$, (c)-(d) $\mu/(v_F|Q|)=0.2$, and (e)-(f) $\mu/(v_F|Q|)=0.4$. 
}
\label{ezz_vs_w}
\end{figure*} 
We now turn our attention to the tilted case when the chemical potential is finite.
In general, the integral in Eq.~(\ref{dispezz})  for $\epsilon_{zz}$ 
is highly
complicated,
and  solutions must  be obtained  numerically. 
Nonetheless, when $|\beta_{s}| < 1$, it is possible to approximate 
$\epsilon_{zz}$ as follows:
\begin{widetext}
\begin{subequations}\label{bigezz}
\begin{align} \label{bigezz'}
\epsilon_{zz}'&=1+\frac{\alpha\mu^2}{\pi\omega^2}\sum\limits_{s=\pm}\frac{1}{\beta_s^3}\Bigg\{
\frac{8}{3}\beta_s-4\arctanh\beta_s
+\ln\bigg|\frac{4\mu^2-\omega^2(1+\beta_s)^2}{4\mu^2-\omega^2(1-\beta_s)^2}\bigg| \nonumber  \\
+\frac{\omega^2}{12\mu^2}&\sum\limits_{t=\pm
  1}\Bigg[t\Big(1+2t\beta_s\Big)\Big(1-t\beta_s\Big)^2\ln\bigg|\frac{4\Gamma^2(1-t\beta_s)^2}{4\mu^2-\omega^2(1-t\beta_s)^2}\bigg| 
-\frac{2\mu}{\omega}
\Big(\frac{4\mu^2}{\omega^2}+ 3-3\beta_s^2)
\ln\bigg| \frac{2\mu-t\omega(1+t\beta_s)}{2\mu+t\omega(1+t\beta_s)}\bigg|\Bigg]\Bigg\},\\
\epsilon_{zz}''&=\frac{\alpha}{6}\sum\limits_{s=\pm}\Theta\Big(
\omega-\frac{2\mu}{1+|\beta_s|}\Big)\Bigg\{1-\frac{1}{2}\Bigg[
1+\frac{3}{2|\beta_s|}\Big( \frac{2\mu}{\omega}-1\Big)\Big( 1-\frac{1}{3\beta^2_s}\Big\{\frac{2\mu}{\omega}-1\Big\}^2\Big)\Bigg] \Theta\Big(
\frac{2\mu}{1-|\beta_s|}-\omega\Big)\Bigg\},  \label{bigezz''}
\end{align}
\end{subequations}
\end{widetext}
where we have decomposed the permittivity into
its real and imaginary components: $\epsilon_{zz}=\epsilon_{zz}' +i\epsilon_{zz}''$.
 If $|\beta_-|=|\beta_+|\equiv |\beta|$, the imaginary part of $\epsilon_{zz}$
 is zero when $\omega<2\mu/(1+|\beta|)$ and increases as a function of
 $\omega$ in the interval
 $2\mu/(1+|\beta|)<\omega<2\mu/(1-|\beta|)$. This expression is
 independent of frequency if $\omega>2\mu/(1-|\beta|)$. 
Next, if we consider
the  limit
 $|4\mu^2-\omega^2|^{1/2}\gg \mu|\beta_s|$, we find that $\epsilon_{zz}$  takes the following form:
\begin{align}
\epsilon_{zz}&=1+\frac{\alpha}{3\pi}\Big[\ln\Big|
\frac{4\Gamma^2}{4\mu^2-\omega^2}\Big|-\frac{4\mu^2}{\omega^2}+i\pi\Theta(\omega-2\mu)\nonumber\\
&-\frac{8\mu^4}{\omega^2}\frac{12\mu^2-5\omega^2}{5(4\mu^2-\omega^2)^2}(\beta_-^2+\beta_+^2) \Big],
\end{align}
where the last term can be viewed as a correction arising from 
the tilt of the Weyl cones. Finally, in the limit $\omega\rightarrow 0$, the off-diagonal component has the form 
\begin{subequations}
\label{gam_approx}
\begin{align}
&\gamma = \frac{\alpha}{\pi \omega}\bigg[2v_FQ -\sum_{s=\pm}\frac{s\mu}{2\beta_s} \bigg(\frac{1}{\beta_s} \ln\bigg|\frac{1+\beta_s}{1-\beta_s}\bigg|-2\bigg)\bigg], |\beta_s|\ll1\\
&\gamma =  \frac{\alpha}{\pi \omega }\sum_{s=\pm}\bigg[\frac{v_FQ}{|\beta_s|}-\frac{s\mu}{2\beta_s}\ln\bigg|\beta_s^2\frac{\Gamma}{\mu}\bigg|\bigg],|\beta_s|\gg1.
\end{align}
\end{subequations}
Thus, for fixed $\beta$, these expressions show that 
 for small values of the tilt parameter,
 $\gamma$ is linear function of $\mu$,
 declining as the chemical potential increases.
 For 
$\beta\gg1$,  $\gamma$ strongly diminishes with  $\mu$, eventually changing sign.
If on the other hand, we have a set chemical potential, increasing  the tilt 
also reduces the gyrotropic effect by weakening $\gamma$, 
and more rapidly  for larger $\mu$. In both regimes, at a vanishing chemical potential $\mu\rightarrow 0$, we recover our previously discussed results
at the charge neutrality point.

\subsection{Susceptibility and epsilon-near-zero responses}\label{subsec1:epsilon} 

When characterizing the nontrivial behavior  of $\rttensor{\epsilon}$ in the WS, there are several relevant 
parameters to consider, including the chemical potential, frequency of the EM wave, tilt of the Weyl cones,
and the node separation parameter $Q$, which is taken to be positive. 
Although it may be possible to generate a mixture of type-I and type-II Weyl points, \cite{boman} we consider 
here the simpler
configuration where
 $\beta_+=-\beta_-=\beta$.
 In presenting the results, we write $\omega$ in units of energy,
 and  the complex component  $\epsilon_{zz}$, is written in terms of its 
 real and imaginary parts:
$\epsilon_{zz} = \epsilon_{zz}'+i\epsilon_{zz}''$.
When presenting results, we plot the normalized susceptibility $\chi_{zz}$,
defined as 
$\chi_{zz}\equiv (\epsilon'_{zz}-1)3\pi/\alpha$. Therefore the ENZ regime corresponds
to  $\chi_{zz}= -3\pi/\alpha$.
We also normalize  
the dissipative component  by $\alpha$.
One of the primary aims is to locate in parameter space,
the particular $\omega$, $\mu$, and $\beta$ that result in the real
part of  $\epsilon_{zz}\sim 0$.
Therefore, when varying the chemical potential in the WS, 
we consider a dimensionless $\mu$ ranging from the charge neutrality point, $\mu=0$,
up to
 $\mu/(v_F |Q|)=0.5$. 
 Similarly, in order to have as complete a picture as possible, a wide spectrum of
dimensionless  frequencies is considered
 corresponding to 
 $0.05 \leq \omega/(v_F |Q|) \leq 1$. When the frequency is not varying,
   we set it to its dimensionless value of $\omega/(v_F |Q|)=0.3$. Of particular importance is the tilt of the Weyl cones, which determines the corresponding $\mu$ and $\omega$ that lead to 
  a vanishing of the real part of $\epsilon_{zz}$. 
 A broad range of cone inclinations covering both
 type-I and type-II scenarios is therefore examined. When computing the integrals, it is necessary to specify an
 energy cutoff $\Gamma$.
Here we 
set $\Gamma/(v_F |Q|) \sim 8$, recalling that the linearized model breaks down 
when $\Gamma>v_F|Q|$, 
and
which is consistent with the requirement 
$\Gamma\gg (\omega, \mu)$ discussed in conjunction with Eq.~(\ref{dispexx}).
Since the linear model is most suitable away from the
Lifshitz transition between type-I and type-II Weyl semimetals, 
a qualitatively correct description
of the system can still be obtained with the linearized
model near the transition by simply
 taking either  larger cut-offs,
or by incorporating higher order momenta into the  linear model.

To begin, in  Fig.~\ref{ezz_vs_w}, we  examine the frequency response of 
the normalized susceptibility $\chi_{zz}$ and $\epsilon_{zz}''$
over a broad range of $\beta$.
The other diagonal components can be compared 
by examining the $\beta=0$ cases, whereby  $\epsilon_{xx,yy}=\epsilon_{zz}$. 
When $\mu=0$, 
we calculate $\epsilon_{zz}$ using  Eq.~(\ref{ezz_ef0}) for $\beta<1$ 
and Eq.~(\ref{ezz2}) for $\beta>1$.
For finite $\mu$ and $\beta<1$, we utilize the expressions in Eq.~(\ref{bigezz'})
and Eq.~(\ref{bigezz''}). 
If however the WS is type-II  with $\beta>1$, we must resort to 
the general integral in Eq.~(\ref{dispezz}) and solve for $\epsilon_{zz}$ numerically.
The results are separated into three columns, where
each  column represents
a different chemical potential, as labeled. Figures \ref{ezz_vs_w}(a)-\ref{ezz_vs_w}(b) show $\chi_{zz}$ and $\epsilon_{zz}$ when $\mu/(v_F|Q|)=0$, \ref{ezz_vs_w}(c)-\ref{ezz_vs_w}(d) $\mu/(v_F|Q|)=0.2$, and \ref{ezz_vs_w}(e)-\ref{ezz_vs_w}(f) when $\mu/(v_F|Q|)=0.4$.  
Beginning with  the charge neutrality point, $\mu/(v_F|Q|)=0$,
we see in Figs. \ref{ezz_vs_w}(a)-\ref{ezz_vs_w}(b) that 
for $\beta\leq1$, $\chi_{zz}$ remains positive over the given frequency range,
similar to a conventional  dielectric. As $\beta$ increases, 
and the system transitions towards a type-II WS ($\beta>1$),
the susceptibility  gets shifted down overall, leading to
regions where $\epsilon_{zz}'<0$. Indeed,  within the type-II regime and $\beta\gg 1$, 
the ENZ frequency can be found from 
Eq.~(\ref{ezz2}), to be approximately 
written as, 
\begin{equation}
\omega^2_{\mathrm{ENZ}}\approx \frac{2\alpha}{\pi}|\beta|\Gamma^2.
\end{equation}
The dissipative component does not depend on frequency, and for $\beta\gg 1$,
declines towards zero.

Next, in Figs. \ref{ezz_vs_w}(c)-\ref{ezz_vs_w}(d), 
the normalized chemical potential is increased to $\mu/(v_F |Q|)=0.2$,
so that now the frequency range of interest is shifted accordingly.
The top panel shows that for the case $\beta=0$, a peak in  $\chi_{zz}$ arises.
This peak emerges due to the inter-band transition, which leads to the singularity arising from the logarithmic term
at $\omega=2\mu$
in Eq.~(\ref{dispexx}). 
For most $\beta$,
we again find at the lowest frequencies, $\epsilon_{zz}'<0$, 
similar to the behavior of some
metals at optical frequencies. 
Upon increasing $\omega$,  $\epsilon_{zz}'$
increases until arriving at the ENZ frequency  where $\epsilon_{zz}'=0$.
The bottom panel (d) exhibits the dissipation characteristics of this WS. 
There are now several distinct features
that $\epsilon_{zz}''$ has
compared to the $\mu=0$ case. In particular, for $\beta=0$,
 the imaginary component abruptly changes from lossless to lossy  at the frequency $\omega=2 \mu$. Increasing 
 $\beta$ causes the dissipation at the transition point to broaden, until $\beta=1$, after which the imaginary component
 becomes independent of the tilt parameter. Note that
 for $\beta<1$,
$\epsilon_{zz}''$ vanishes
for $\omega\lesssim2\mu/(1+\beta)$, and 
 increases  as a function of
 $\omega$ in the interval
 $2\mu/(1+\beta)<\omega<2\mu/(1-\beta)$ [See Eq.~(\ref{bigezz''})]. Lastly, in panels (c) and (d), a larger 
 chemical potential corresponding to the dimensionless value of $\mu/(v_F |Q|)=0.4$ is considered. It is observed that
 when increasing $\mu$, there is a widening of the frequency window in which the ENZ response occurs. 
 There is also a broadening off the imaginary component resulting in finite dissipation over more 
 frequencies.

\begin{figure}[!tbp]
\centering
\includegraphics[width=0.45\textwidth]{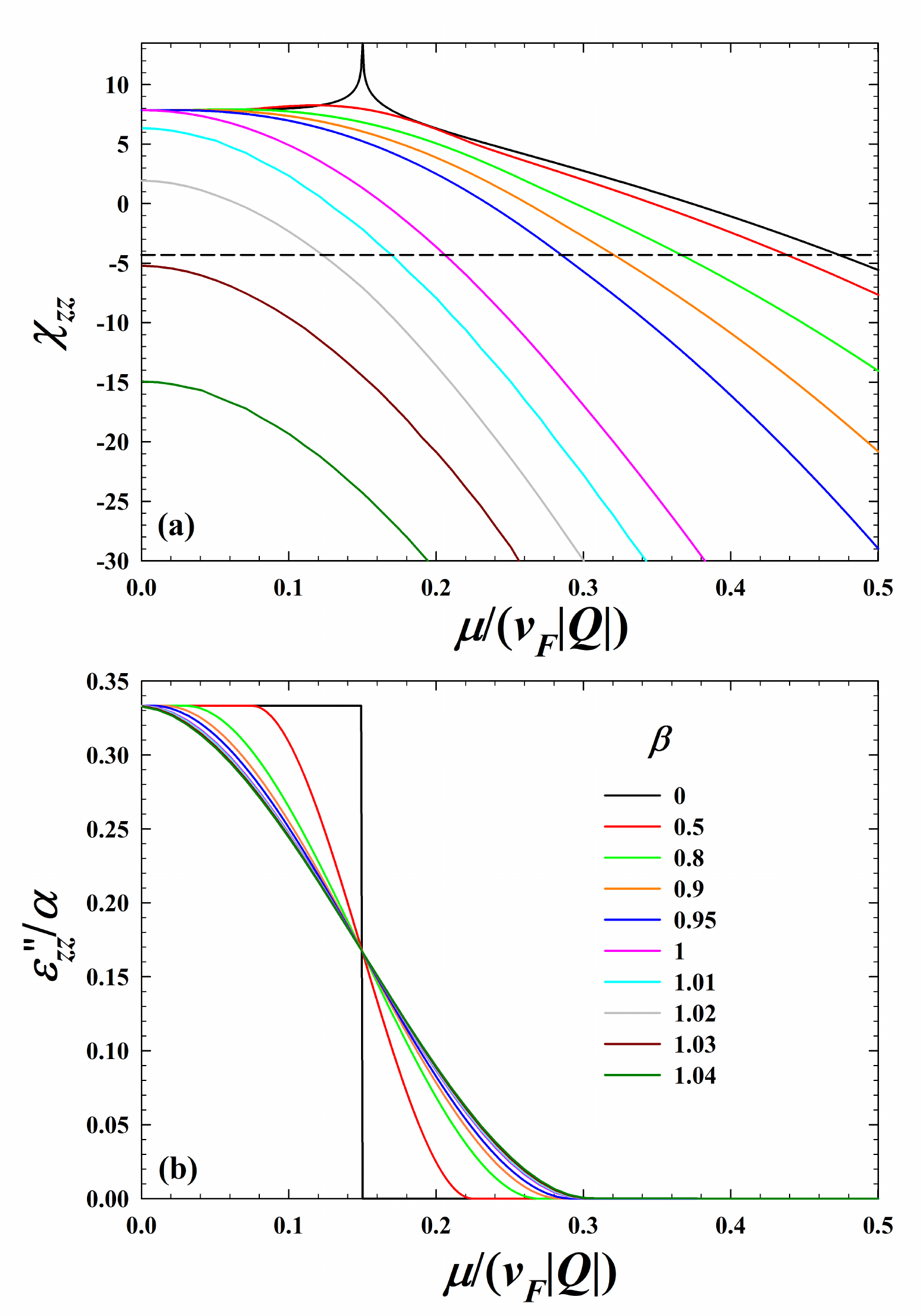}
\caption{(Color online). 
(a)
Illustration of 
how the  
susceptibility ${\chi}_{zz}$
is affected by changing 
 the chemical potential in a Weyl semimetal. 
 Increasing the tilt of the Weyl cones
is seen to require a smaller $\mu$ for an ENZ response,
corresponding to the intersection of the curves with the dashed line.
(b) The normalized dissipative response $\epsilon_{zz}''/\alpha$.
In both case, the normalized frequency corresponds to $\omega/(v_F |Q|)=0.3$.
}
\label{ezz_vs_ef}
\end{figure} 
One of the salient features of Weyl semimetals is the ability to systematically change   their
chemical potential.
It is possible to  shift $\mu$ about the charge neutrality point through
doping, 
varying the
temperature,  
or altering the lattice constant of the material through pressure variations.\cite{chin,qiu1,hau}
For instance, upon increasing the temperature, 
the number of thermally excited charged carriers 
increases near the Weyl points that increases the chemical potential\cite{chin}. 
The injection of various dopants into the Weyl semimetal also can increase 
the number of free charged carriers, depending on the dopant type \cite{hau,tim,wang}. To 
examine how changes in  $\mu$ 
can alter the EM response of a WS, 
we examine 
in Fig.~\ref{ezz_vs_ef}(a) $\chi_{zz}$, and \ref{ezz_vs_ef}(b) the dimensionless
$\epsilon_{zz}''$ as functions of $\mu$. 
Figure \ref{ezz_vs_ef}(a) shows that for  $\beta\leq1$ and $\mu=0$, all curves originate at 
$\chi_{zz} = \ln (4\Gamma^2/\omega^2)\approx 7.85$
[see Eq.~(\ref{ezz_ef0})].
For  Weyl cones that
are  tilted with $\beta\leq1$,
increasing $\mu$ causes a splitting of the curves
which then monotonically  decline. 
Further increases in $\beta$ causes   
$\chi_{zz}$ to shift downward, 
 becoming negative for all $\mu$. 
Figure \ref{ezz_vs_ef}(b) exhibits  how the normalized 
dissipation  $\epsilon_{zz}''/\alpha$  can be 
drastically manipulated through changes in $\mu$.
The type-I case at $\mu=\omega/2$ has an abrupt transition 
 at $\beta=0$. By increasing the tilt angle of the Weyl cones, 
the dissipative response broadens and 
$\epsilon_{zz}''$ is finite over a larger interval of chemical potentials.
This can be of significance in anisotropic ENZ systems with dielectric losses
and 
system parameters are tuned to 
control beam directivity\cite{leaky} or
EM wave absorption. \cite{feng, hmm}

\begin{figure}[!tbp]
\centering
\includegraphics[width=0.45\textwidth]{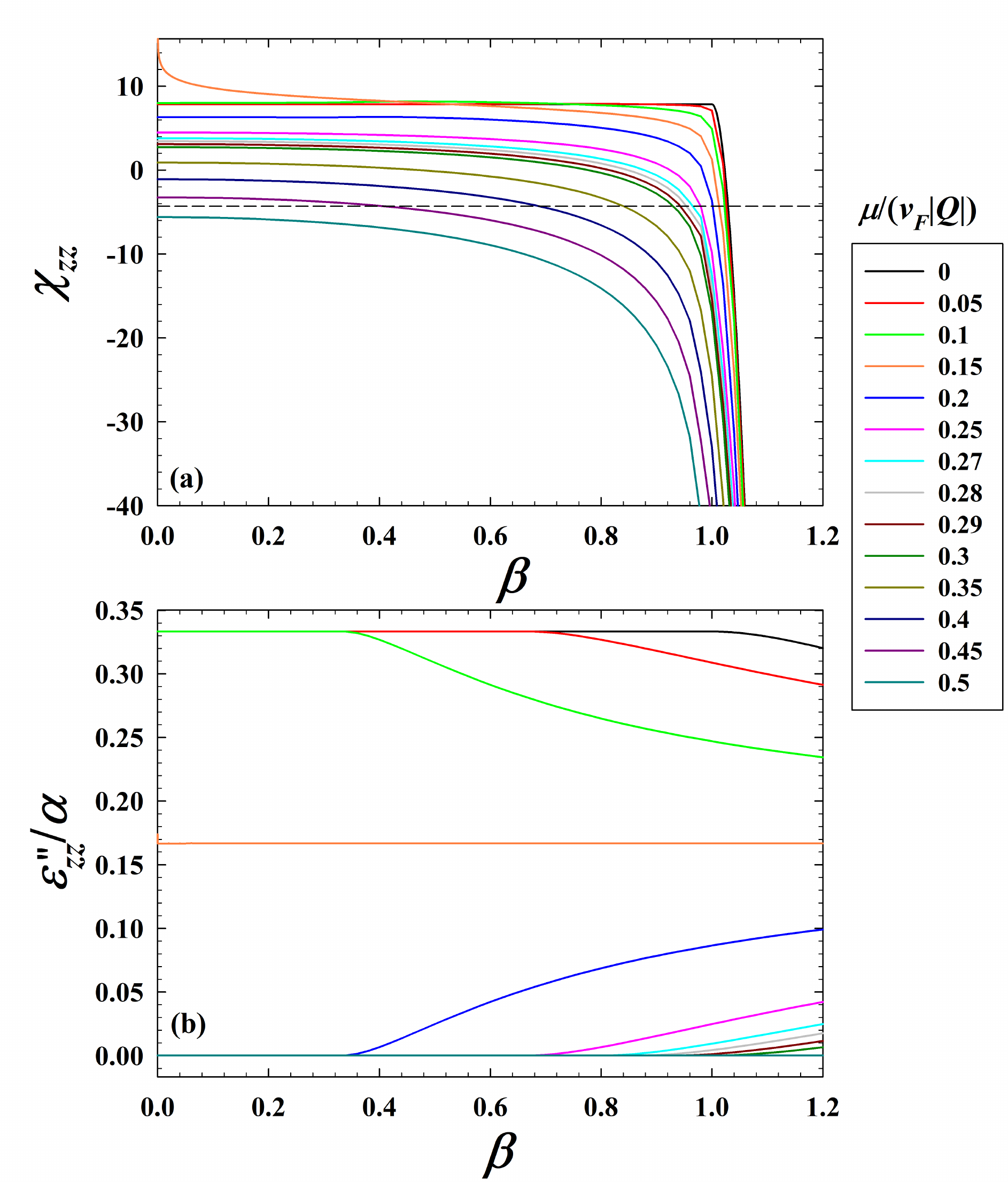}
\caption{(Color online). 
Generation of an ENZ state through tilting of the Weyl cones:
Through variation in $\beta$, panel (a) shows that ${\chi}_{zz}$ can transition
to an ENZ state as represented by the dashed horizontal line.
Panel (b) displays the imaginary component $\epsilon_{zz}''/\alpha$,
where it is seen that $\beta$ strongly affects the  dissipation nature of the WS. 
The same $\mu$ values
 are used  in both panels.}
\label{ezz_vs_beta}
\end{figure} 
Next, in Fig.~\ref{ezz_vs_beta} we investigate
how  tilting of the  Weyl cones influences 
the normalized susceptibility, \ref{ezz_vs_beta}(a), and 
 dissipation \ref{ezz_vs_beta}(b). Each curve represents
 different equally spaced $\mu$,  normalized as shown  in the legend.
Starting with  $\mu=0$ and $\beta<1$, we observe that 
the permittivity is independent  of the tilt, in agreement with Eq.~(\ref{ezz_ef0}).
As $\beta$ increases however, $\chi_{zz}$ rapidly drops.
The same type of behavior is seen in the other curves with small $\mu$,
where regions of relatively constant $\chi_{zz}$ diminish as $\mu$ increases.
Thus, when the chemical potential is in the vicinity of the charge neutrality point, 
only a  type-II WS, with $\beta>1$, can exhibit ENZ behavior.
As $\mu$ increases, each corresponding  curve gets shifted down towards the
 ENZ line, so that smaller $\beta$
 can lead to a transition to a metallic-like state.
Eventually, for the largest chemical potential shown, 
$\chi_{zz}$ cannot exhibit an ENZ response for any value of tilt.
The observed peak at
$\omega/\mu=2$ (corresponding to $\mu/(v_F |Q|)=0.15$) becomes diminished for other values of $\mu$.
For  relatively weak chemical potentials,  $\mu/(v_F |Q|) \leq 0.2$, 
an ENZ response is seen to
be induced only when the WS is type-II. 
For $\beta\leq1$, the ENZ regime arises only for
larger $\mu$. 
For example, 
the ENZ state is reached at $\beta\approx0.4$, and
$\mu/(v_F |Q|) =  0.45$.
Thus, if the Weyl semimetal 
is to demonstrate an ENZ response,
it should be 
type-I with sufficiently  large $\mu$,
or it can be type-II with smaller $\mu$. 
In either case, the dissipative component will be strongly affected, as
panel (b) illustrates how cone tilt inclinations strongly influence  
$\epsilon_{zz}''$. 
The normalized $\epsilon_{zz}''$ component is shown to not exceed $1/3$. 
When $\mu/(v_F|Q|)=0.15$ (or equivalently $\omega/\mu=2$),
 the dimensionless $\epsilon_{zz}''$ is constant and has the value $1/6$.
 Above this value of the chemical potential, 
 the dissipative response  tends to decline as $\beta$ increases,
 while below it, the dissipation increases as $\beta$ increases.
 At the charge neutrality point $\epsilon_{zz}''$, 
 is unaffected by changes in the tilt for type-I, but for type-II ($\beta\gtrsim1$), 
there is a weak decline, according to   Eq.~(\ref{ezz2}).
Comparing with (a), it is evident that for $\mu/(v_F|Q|)\gtrsim 0.3$, an ENZ
response with zero effective loss can be achieved for a type-I WS.
This situation could be relevant to waveguide structures, where   
 localized electromagnetic  waves propagate over long distances near the surface of
the WS.

To estimate the feasibility of achieving an ENZ response in a type-II WS,
we consider\cite{ma,wyl1} 
$Q=1\,{\rm nm}^{-1}$, $\mu=0.08\,{\rm eV}$,
 $\beta=1.02$, and $\omega=0.2\,{\rm eV}$,  
 to give $\epsilon'_{zz}\sim 0$, and $\epsilon''_{zz}\sim0.5$. 
Although this estimate corresponds to a pair of points,   
to relate to a material such as  $\rm TaAs$, which  has $12$ pairs of Weyl cones, including  8 nodes with Fermi energy 
 $\mu\approx 1\,{\rm meV}$, and 16 nodes with $\mu\approx 15\,{\rm meV}$,
  $\alpha$ should be multiplied by 4 and 8 respectively, to account for the additional nodes.  

\section{Perfect Absorption in Weyl Semimetal Structures}\label{sec2:abs_rfl}
In this section, we make use of the results for ${\rttensor{\epsilon}}$
presented in the previous section to demonstrate how a WS structure 
in the ENZ regime can be tuned to
exhibit perfect absorption of EM waves over a broad range of incident angles
and system parameters, thus revealing 
a practical platform for the control of EM radiation.

\subsection{Maxwell's equations and theory}\label{subsec2:max}

\begin{figure}[!tbp] 
\centering
\includegraphics[width=0.45\textwidth]{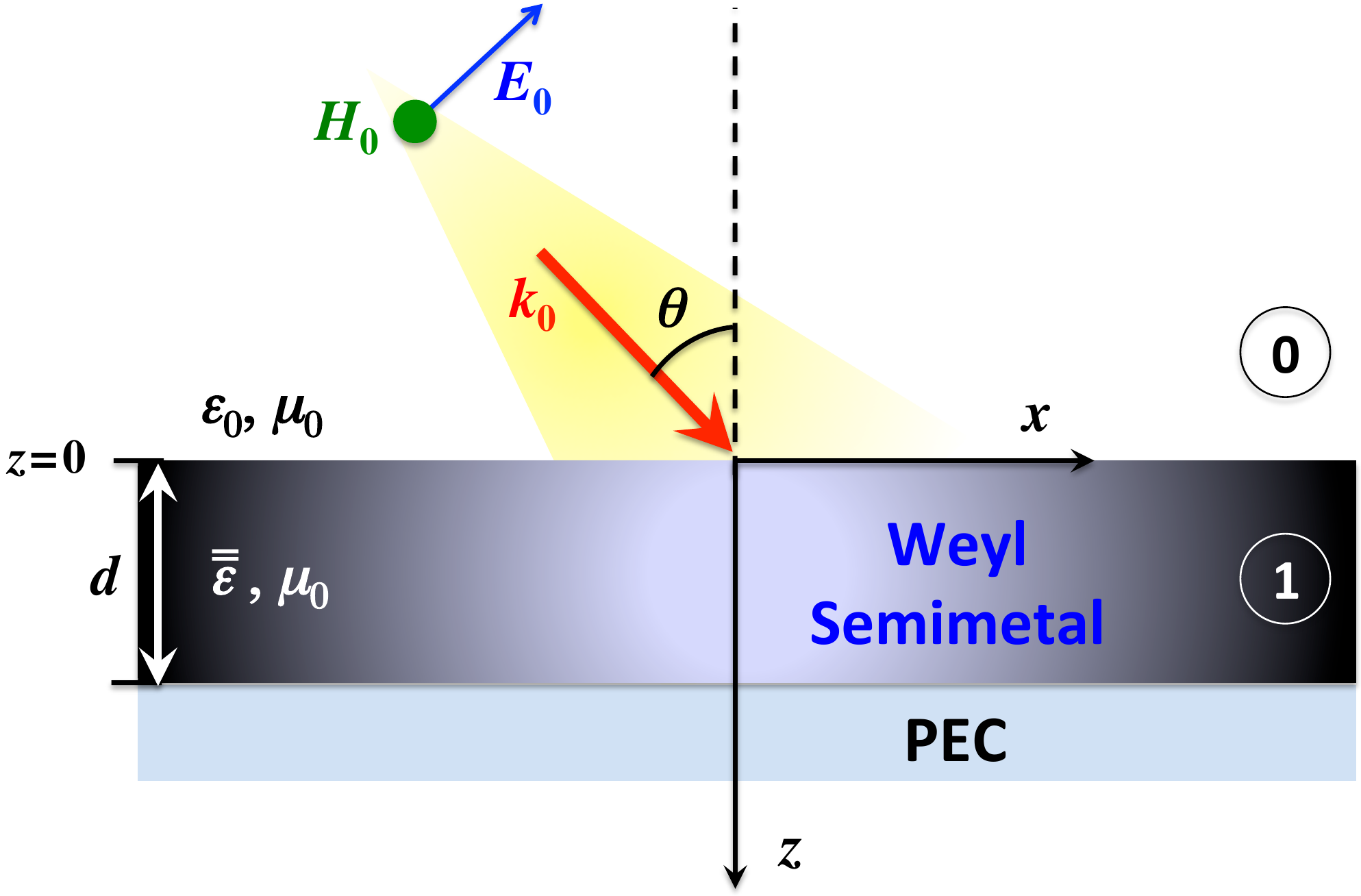}
\caption{(Color online).  Schematic of the configuration
  involving a Weyl semimetal in region $\circled{1}$ with width $d$
  atop 
a perfectly conducting substrate. 
The Weyl
semimetal layer is exposed to an electromagnetic wave from the vacuum
region $\circled{0}$. The incident electric field is polarized in the $\var{x-z}$ plane,
and the magnetic field is polarized along $y$.
The incident wavevector ${\bm k_0}$ makes an angle $\theta$ with the $z$ axis.
The separation of the Weyl nodes is taken to be along the $z$-axis}
\label{diagramxz}
\end{figure}

We investigate the reflection and absorption
of EM waves from 
the layered configuration shown in 
 Fig.~\ref{diagramxz}, which 
 consists of a planar
 Weyl semimetal (region $\circled{1}$) adjacent to 
  a metallic substrate with perfect conductivity (PEC).
  The electric field of the  incident wave is polarized in the $\var{x-z}$ plane,
  so that 
  the permittivity component
$\epsilon_{zz}$ plays a significant role in
 the overall EM response.  
The plane wave is 
incident from vacuum (region $\circled{0}$)
  with wavevector ${\bm k}_0$ also in the $\var{x-z}$ plane:
${\bm k}_0=\hat{\bm x}k_{0x} + \hat{\bm z} k_{0z}$. 
If the incident wave was propagating  in the $\var{x-y}$ plane,
the TE and  TM modes would become decoupled, and
the reflectivity characteristics for the TM modes would  not 
depend on the $\epsilon_{zz}$ component
of the permittivity tensor.
For propagation in the $\var{x-z}$ plane,
the TE and TM modes can no longer be separated, and 
the EM response of the WS structure is governed mainly by
$\epsilon_{zz}$.

The incident electric and magnetic  fields 
 thus have the following forms: 
${\bm E} =(E_{x0} \hat{\bm x}+E_{z0}\hat{\bm z}) e^{i(k_{0x} x +k_{0z} z-\omega t)}$,
and ${\bm H} =H_{y0} \hat{\bm y} e^{i(k_{0x} x +k_{0z} z-\omega t)}$. Here
$k_{0 x}$ is invariant across each layer, with 
$k_{0 x} = k_0 \sin\theta$,  $k_{0 z} = k_0 \cos \theta$, and $k_0=\omega/c$.
For both regions $\circled{0}$ and $\circled{1}$, we  implement   
Maxwell's equations for time harmonic fields,
\begin{subequations}
\begin{align} {\bm \nabla} \times {\bm E}_i &= i \omega \mu_0 {\bm H}_i, \label{h1} \\
{\bm  \nabla} \times {\bm H}_i &= -i\omega {\bm D}_i,
  \label{d1}
\end{align}
\end{subequations}
where $i=0$ or $1$.
Within the WS, the propagation vector ${\bm k}_1$ replaces the spatial derivatives,
transforming  Maxwell's equations into
 the forms, ${\bm k}_1 \times {\bm E}_1= \omega \mu_0 {\bm H}_1$ and
${\bm k}_1 \times {\bm H}_1 = -\omega {\rttensor{\epsilon}} \epsilon_0  {\bm E}_1$. 
These two equations together result in the following  expression for the ${\bm E}_1$ 
field in $\bm k$-space:
\begin{align}\label{wave}
{\bm k}_1 \times ({\bm k}_1 \times {\bm E}_1) = -k_0^2  {\rttensor{\epsilon}}  {\bm E}_1. 
\end{align}
 Using ${\bm k}_1 = k_{0x} \hat{\bm x}+k_{1z} \hat{\bm z}$, and the identity  
${\bm k}_1 \times ({\bm k}_1 \times {\bm E}_1) = {\bm k}_1({\bm k}_1 {\bm E}_1) - k_1^2 {\bm E}_1 $,
permits expansion of  Eq.~(\ref{wave}),
\begin{align}\label{dmatrix}
\begin{pmatrix} 
k_0^2\epsilon_{xx}-k_{1z}^2  &i k_0^2\gamma &k_{1z} k_{0x} \\
i k_0^2 \gamma&k_{\perp}^2- k^2_{0}\epsilon_{yy}& 0 \\
k_{0x} k_{1z}  & 0  &k_0^2\epsilon_{zz}-k_{0x}^2
\end{pmatrix} \hspace{-.2cm}
\begin{pmatrix}
E_{x1} \\
E_{y1} \\
E_{z1} 
\end{pmatrix}= 0,
\end{align}
where $k_{\perp}=\sqrt{k_{1z}^2+k_{0x}^2}$, and $k_{1x}=k_{0x}$
due to translational invariance. 
The coupling of all three components of the $\bm E$ fields in 
Eq.~(\ref{dmatrix}) illustrates that although the electric field of the incident beam  is polarized in the 
$\var{x-z}$ plane,
it can now acquire an additional $y$ component when entering the gyrotropic medium. Similarly,
despite having an initial polarization state along the $x$ direction, the incident ${\bm H}$ field can also 
in general become polarized in all three directions once entering the WS.
Thus, the EM wave exiting the WS structure can have a
different overall polarization state  that depends on
the WS material and geometrical parameters.

\subsection{Results and discussions}\label{subsec2:results}

Since the incident beam 
propagates in the $\var{x-z}$ plane 
with wavevector 
${\bm k}=\hat{\bm x}k_{0x} + \hat{\bm z} k_{0z}$ (see Fig.~\ref{diagramxz}),
each component of $\rttensor{\epsilon}$ must be accounted for in the
EM response of the WS. 
Taking the determinant of the matrix in
Eq.~(\ref{dmatrix}) 
and setting it equal to zero,
gives the dispersion equation for the WS that can be solved for $k_{1z}$:
\begin{align} \label{disp2p}
(\epsilon_{xx} k_0^2-k_{\perp}^2)(\epsilon_{xx}\epsilon_{zz} k_0^2&-
\epsilon_{xx} k_{0x}^2 -\epsilon_{zz} k_{1z}^2)\nonumber \\
&+k_0^2(k_{0x}^2-\epsilon_{zz} k_0^2)\gamma^2=0.
\end{align}
Solving for the roots in Eq.~(\ref{disp2p})  results in two types of solutions to $k_{1z}$,
 denoted by $ k_+$ and $k_-$. We have,
\begin{align}
 k_{\pm}^2 = &\frac{k_0^2}{2\epsilon_{zz}}\Biggl[
2  \epsilon_{zz}\epsilon_\parallel-(\epsilon_\parallel+\epsilon_{zz})
\sin^2\theta \nonumber\\
&\pm \sqrt{4\epsilon^2_{zz} \gamma^2-
4\epsilon_{zz}\gamma^2\sin^2\theta+
(\epsilon_\parallel-\epsilon_{zz})^2 \sin^4\theta}\Biggr],
\end{align}
where $\epsilon_\parallel$ represents  
the components of  ${\rttensor{\epsilon}}$ parallel to the interfaces:
$\epsilon_{xx}=\epsilon_{yy}=\epsilon_\parallel$.
The dispersion equation (\ref{disp2p}) 
can now be compactly written in terms of the two types of waves:
\begin{align}
k_0^2 \epsilon_{zz}(k_{1z}^2-k_+^2)(k_{1z}^2-k_-^2)=0.
\end{align}

For the configuration shown in Fig.~\ref{diagramxz}, where the $\var{x-y}$ plane is
translationally invariant,  the magnetic field components 
in the vacuum region, ${\bm H}_0$,
are written in terms of
 incident and reflected waves:
\begin{subequations}
\begin{align} 
H_{x0} &= r_3 e^{-i k_{0z} z} e^{i k_{0x} x}, \label{hobag1} \\
H_{y0} &= (e^{i k_{0z} z} + r_1 e^{-i k_{0z} z})e^{i k_{0x} x},\label{hobag2}\\
H_{z0} &= r_2 e^{-i k_{0z} z} e^{i k_{0x} x},\label{hobag3}
\end{align}
\end{subequations}
where the $x$ and $z$ components represent
the change of the incident polarization state upon reflecting from the Weyl semimetal. 
The coefficients $r_2$ and $r_3$ 
take into account the generation of additional
polarization components 
upon interacting with the gyrotropic WS layer.
Note that from the Maxwell's equation 
${\bm\nabla\cdot}{\bm H}_0=0$,
there exists a simple relation between the coefficients $r_2$ and $r_3$:
\begin{align}
r_3=\frac{k_{0z}}{k_{0x}} r_2.
\end{align}
From  the magnetic field components above, we can use Eq.~(\ref{d1}) 
to easily deduce
the electric field components for region $\circled{0}$.

For region $\circled{1}$, when using Maxwell's equations,
we need to take into account the anisotropic nature of the WS.
The general solution to the ${\bm E}$ field in the WS region is
thus
 a linear combination
of the four wavevector components  $k_{1z}=\{k_+,-k_+,k_-,-k_-\}$:
\begin{align} \label{ey1}
E_{y1}=(a_1 e^{i k_+ z}+a_2e^{-i k_+ z}+a_3 e^{i k_- z}+a_4 e^{-i k_- z})e^{i k_{0x} x}.
\end{align}
To determine the coefficients $\{a_1,a_2,a_3,a_4\}$, it is necessary to invoke 
matching interface conditions and  
boundary conditions.
But first we must construct the remaining $\bm E$ and $\bm H$ fields.
This is achieved via the two Maxwell's equations, Eqs.~(\ref{h1}) and (\ref{d1}). 
First, using (\ref{d1}) gives the following relations:
\begin{subequations}
\begin{align}
&\frac{\partial H_{y1}}{\partial z}=i\omega \epsilon_0(\epsilon_\parallel E_{x1}+i\gamma E_{y1}),\\
&\frac{\partial H_{x1}}{\partial z}-ik_{0x} H_{z1}=i\omega \epsilon_0 (i\gamma E_{x1}-\epsilon_\parallel E_{y1}), \label{emid} \\
&k_{0x} H_{y1}=-\omega \epsilon_0 \epsilon_{zz} E_{z1}.
\end{align}
\end{subequations}
While, from Eq.~(\ref{h1}) we have,
\begin{subequations}
\begin{align}
&\frac{\partial E_{y1}}{\partial z}=-i\omega\mu_0 H_{x1}, \label{hx1} \\
&\frac{\partial E_{x1}}{\partial z}-ik_{0x} E_{z1}=i\omega\mu_0 H_{y1}, \label{hmid}\\
&k_{0x} E_{y1}=\omega \mu_0 H_{z1}, \label {hz1}
\end{align}
\end{subequations}
where we have used the fact that $x$ component is invariant, i.e., $\partial_x \rightarrow i k_{0x}$.
Inserting Eq.~(\ref{ey1}), into the equations above,
it is now possible to write all components of the EM field in terms of the coefficients $\{a_1,a_2,a_3,a_4\}$.
For example, $H_{x1}$ and $H_{z1}$ are easily found from Eqs.~(\ref{hx1}) and (\ref{hz1}) respectively. From that,
one can solve Eq.~(\ref{emid}) for
$E_{x1}$, and so on.

Upon matching the tangential electric and magnetic fields at the vacuum/WS interface, and 
using the boundary conditions of vanishing tangential electric fields at the ground plane,
it is straightforward to determine the unknown coefficients. The first reflection coefficient $r_1$
is defined as $r_1=1-r_0$, where
\begin{widetext}
\begin{gather}
r_0=\frac{2 k_z^2[k_+ q^2_- \cos(k_+ d) \sin(k_- d)-[k_- q^2_+\cos(k_- d)
+i k_{0z} (k_+^2-k_-^2) \sin(k_- d)]\sin(k_+d)]}
{k_-\cos(k_- d)[f_2 \sin(k_+d)+i\epsilon_{zz} k_{0z} k_+ (k_+^2-k_-^2) \cos(k_+ d)]
+\sin(k_-d)[f_1 k_+ \cos(k_+d)-ik_{0z} k_z^2 (k_+^2-k_-^2) \sin(k_+d)]},\\\nonumber
k_z=k_0\sqrt{\epsilon_{zz}-\sin^2\theta}, \;\;\;\; q_{\pm}=\sqrt{\epsilon_\parallel k_0^2 -k_{0x}^2-k_\pm^2},\\\nonumber
f_1=k_z^2 q_-^2-\epsilon_{zz} k_{0z}^2 q_+^2, \;\;\;\;f_2=-k_z^2 q_+^2+\epsilon_{zz} k_{0z}^2 q_-^2.
\end{gather}
\end{widetext}
The $r_2$ coefficient is expressed compactly in terms of $r_1$ and $r_0$:
\begin{align}
r_2=\frac{k_{0x} \gamma[k_z^2(r_1+1)\sin(k_-d)-i\epsilon_{zz} k_{0z} k_- r_0\cos(k_- d)]}
{\epsilon_{zz} q_+^2[i k_{0z} \sin(k_- d)-k_- \cos(k_- d)]}.
\end{align}

\begin{figure}[!thb] 
\centering
\includegraphics[width=0.45\textwidth]{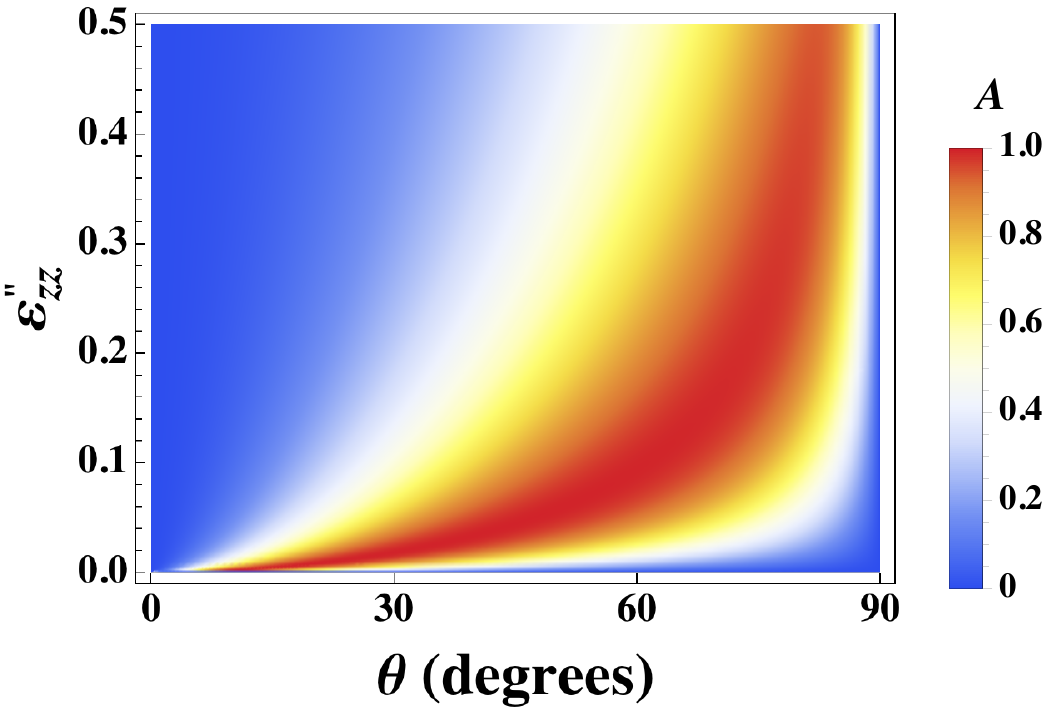}
\caption{(Color online). 
Color map demonstrating how the absorptance $A$
varies as a function of the incident angle $\theta$ and
dissipation $\epsilon''_{zz}$. 
The system is in the ENZ regime, whereby $\epsilon'_{zz}=0$. 
The normalized frequency is set to $\omega/(v_F |Q|)=0.3$,
and $d/\lambda=1/100$.
}
\label{2dtheta}
\end{figure} 

In the absence of gyrotropy,  $\gamma=0$, $r_2=r_3=0$,
and
the corresponding  reflection coefficient $r_1$ reduces to
\begin{align} \label{r1_gam0}
r_1=1-\frac{2 k_{-}}{k_{-} + i k_{0z}  \epsilon_\parallel \cot(k_{-}d)},
\end{align}
in which 
$k_-=\sqrt{\epsilon_\parallel(k_0^2-k_{0x}^2/\epsilon_{zz})}$.
Thus, when  $\gamma=0$,
the reflection coefficient reverts to that of  a diagonally anisotropic
medium \cite{hmm}, as it should.
When the
gyrotropic parameter vanishes, 
the incident electric field that is polarized in the $\var{x-z}$ plane, remains in that plane after
interacting with the WS.

In determining the absorptance $A$ of the WS system,
it is beneficial to study the energy flow in the vacuum region.
To this end, we
consider the time-averaged Poynting vector in the direction 
perpendicular to the interfaces (the $z$ direction), $S_{z0}
=\Re{\{E_{x0} H_{y0}^*-E_{y0} H_{x0}^*\}}/2$.
Inserting the electric and magnetic fields
calculated for region $\circled{0}$ above, we find,
\begin{align} \label{bigA}
A= 1-\left | r_1 \right |^2- \left| r_2 \right |^2-\left| r_3 \right |^2.
\end{align}
Here $A$ is defined as $S_{z0}/S_0$, where $S_0\equiv k_{0z}/(2 \epsilon_0 \omega)$
 is the time-averaged Poynting vector for a
plane wave traveling 
in the $z$ direction.

\begin{figure}[!tbp] 
\centering
\hspace*{-1.5cm} \includegraphics[width=0.4\textwidth]{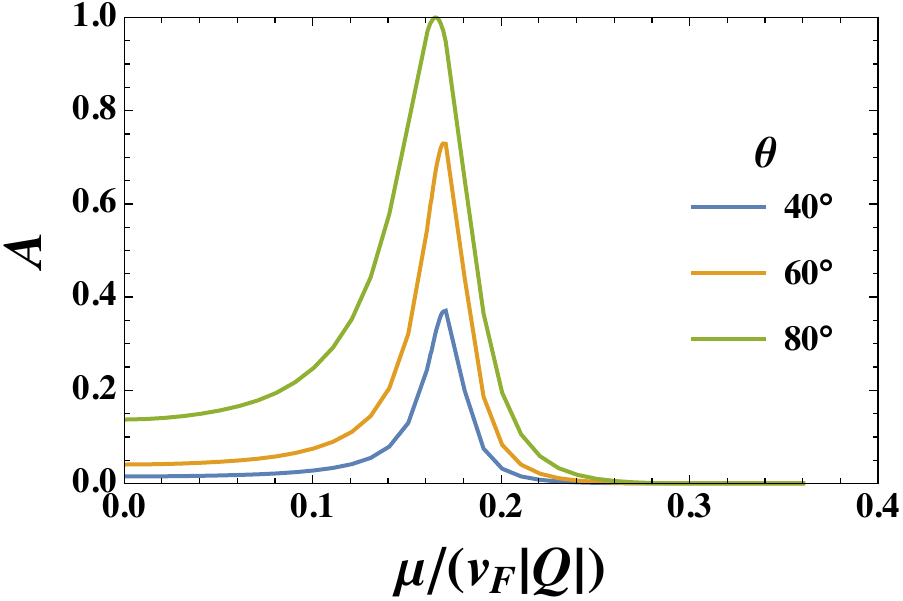}
\caption{(Color online). 
The absorptance $A$ 
as a function of normalized chemical potential  $\mu/(v_F|Q|)$
for a type-II WS structure having an ENZ response.
Three different orientations $\theta$ of the incident beam are shown.
The tilting parameter 
corresponds to
$\beta =1.01$, 
the normalized frequency is  $\omega/(v_F |Q|)=0.3$,
and $d/\lambda=1/100$.
}
\label{AvEF}
\end{figure} 
Having established the methods for determining the absorption and reflection coefficients, 
we now consider a range of material and geometrical parameters
that lead to perfect absorption in
the ENZ regime where
$\epsilon'_{zz}\approx 0$.
The dissipative  component $\epsilon''_{zz}$ on the other hand can vary, as it
plays a crucial role in how electromagnetic energy is absorbed by the system.
For a given WS width $d$, frequency  $\omega$ of the incident wave,
and  orientation $\theta$,
the absorption [Eq.~(\ref{bigA})] can be  calculated by
incorporating  the results of Sec.~\ref{sec1:method},
which gives the various  
 $\mu$ and $\beta$
 that lead to  an ENZ response,
 and allows the remaining   
components of the tensor $\rttensor{\epsilon}$ to be determined 
via  Eqs.~(\ref{dispexx})-(\ref{dispgam}).

\begin{figure}[!tbp] 
\centering
\includegraphics[width=0.48\textwidth]{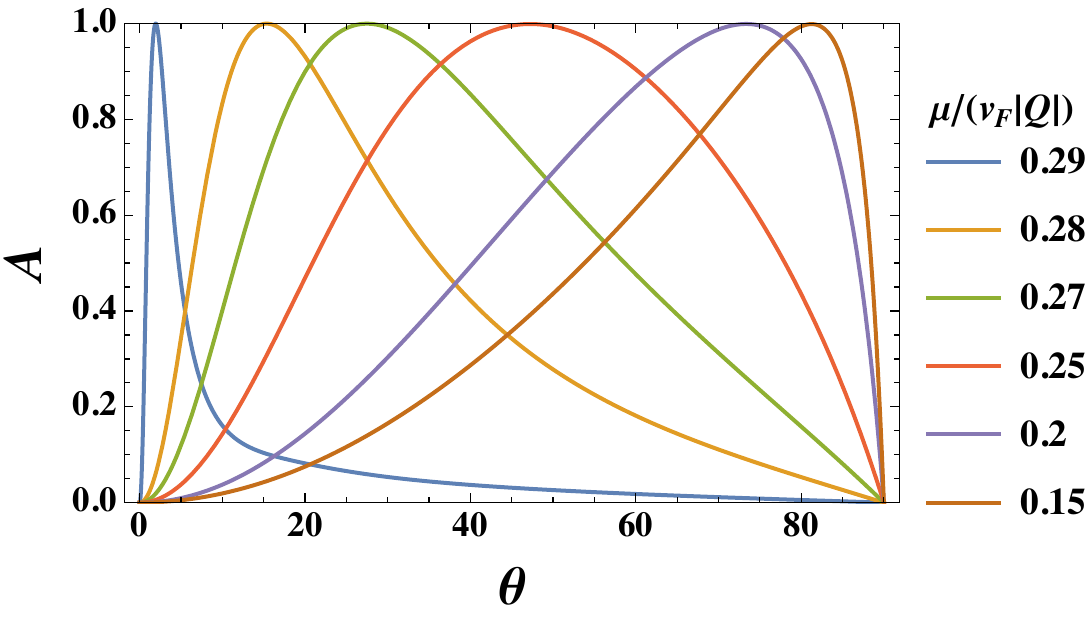}
\caption{(Color online). 
The absorptance $A$ as a function of incident angle $\theta$. 
A broad range of  normalized chemical potentials are considered (see legend).
For
$\mu/(v_F|Q|) =0.29,0.28,0.27,0.25$, and $0.2$, the Weyl semimetal  should be  type-I, with
$\beta\sim0.94,0.96,0.97,0.98$, and $0.99$, respectively. 
To achieve perfect absorption for larger
incident angles that more closely approach grazing ($\theta\rightarrow 90^\circ$),
the $\mu/(v_F|Q|) =0.15$ case requires a type-II WS,
with $\beta\sim 1.01$. 
In all cases the normalized frequency is set to $\omega/(v_F |Q|)=0.3$,
and $d/\lambda=1/100$.
}
\label{Avtheta}
\end{figure}

The results from Sec.~\ref{sec1:method}
offer clear   
 guides 
for identifying  
 ENZ regions of the parameter space. 
  For example, it was observed in Fig.~\ref{ezz_vs_w},
that for the  range of frequencies considered, 
it is necessary for the chemical potential 
to be nonzero for the dissipative component  $\epsilon''_{zz}$
to have significant variations.
We show below that 
$\epsilon''_{zz}$ 
plays a crucial role
in determining  how much of the incident beam
is perfectly absorbed\cite{feng}, and that
 strong absorption can arise over a broader range of $\theta$ 
not only when $\epsilon''_{zz}$ corresponds to moderate loss, 
but also when it is very small.
Thus, we focus on the more interesting cases 
when   $\mu$ is away   from the charge neutrality point ($\mu=0$).
Indeed, Figs.~\ref{ezz_vs_w}(c)-\ref{ezz_vs_w}(f) showed that by having a
finite $\mu$, $\epsilon''_{zz}$ can be continuously tuned
from 
zero to a situation having appreciable dissipation.
If the dissipation vanishes entirely, the type of  perfect absorption
studied here does not arise
since all incoming waves are reflected back from the ground plane. 
Increasing the loss makes
it possible  at appropriate frequencies and orientations of  the incident beam,
for the waves in the WS to destructively interact and ultimately dissipate through Joule heating. 
For most of the frequencies of interest here, which satisfy  $ \omega/(v_F |Q|)<2\mu/(v_F|Q|)$,
the component of the permittivity parallel to the interfaces $\epsilon_\parallel$ [Eq.~(\ref{dispexx})], 
is purely real, as it
has no interband contribution to the optical
conductivity.

\begin{figure*}[!tbp] 
\centering
\includegraphics[width=0.45\textwidth]{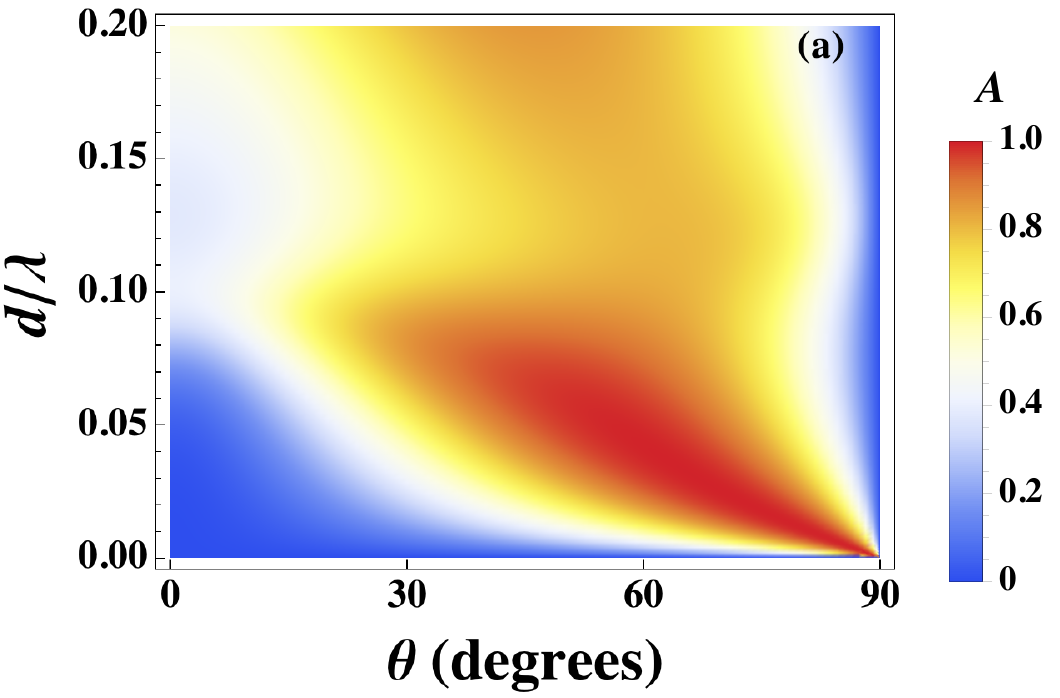}
\includegraphics[width=0.45\textwidth]{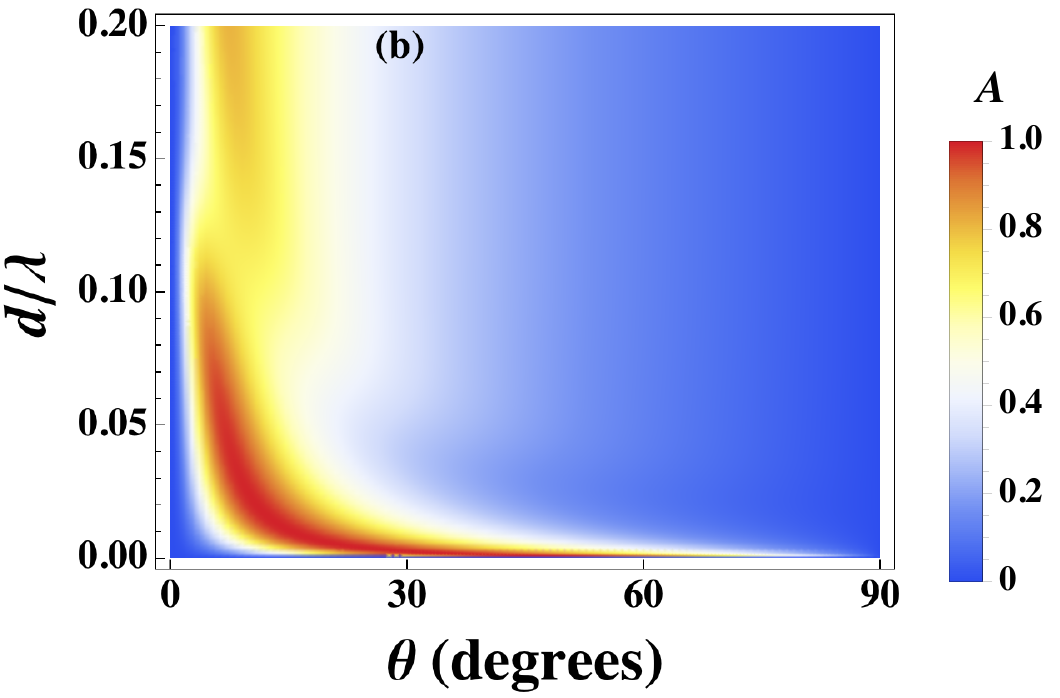}
\caption{(Color online). 
Color maps illustrating the perfect absorption regions
 for various normalized widths $d/\lambda$ and incident angles $\theta$.
 In (a) $\beta\sim1.01$, and $\mu/(v_F |Q|)=0.15$,
 In panel (b) $\beta\sim0.96$,  and 
 $\mu/(v_F |Q|)=0.28$.  
 In both panels the normalized frequency is set to $\omega/(v_F |Q|)=0.3$,
}
\label{Avtheta2}
\end{figure*}

To illustrate how $\epsilon''_{zz}$ directly impacts the absorption characteristics 
of the WS structure shown in Fig.~\ref{diagramxz},   
we present in Fig.~\ref{2dtheta},
a color map that depicts  the absorptance $A$
as a function of
 the incident angle $\theta$ and the loss $\epsilon''_{zz}$.
 We consider the scenario where
 the WS is assumed to have an ENZ response, $\epsilon'_{zz}=0$, so that 
$\epsilon_{zz}$ is described entirely by its imaginary component. 
As was extensively discussed  in Sec.~\ref{sec1:method},  $\epsilon''_{zz}$ generally depends on
 several  system parameters.
Here however, 
in order to isolate the effects of dissipation and its relation to 
angles of high absorption, $\epsilon''_{zz}$  varies independently.
In this example, we have  $\mu/(v_F |Q|)=0.2$, and $\gamma=3$. 
As Fig.~\ref{2dtheta} shows, 
depending on $\epsilon''_{zz}$,
perfect absorption can be achieved for
virtually any incident angle.
Thus, for example,
to determine the 
WS parameters 
needed to achieve perfect absorption around normal incidence
($\theta\sim0^\circ$), 
it is  only necessary to identify configurations 
where 
$\epsilon'_{zz}\approx0$ for small $\epsilon''_{zz}$.
The fact that for angles near normal incidence, the
loss must be extremely weak for perfect absorption is consistent with
the response of isotropic ENZ slabs\cite{enghetta}. 
On the other hand, to have the entire incident wave's energy absorbed 
at broader angles ($\theta\sim90^\circ$),
the ENZ structure must exhibit greater dissipation, with  $\epsilon''_{zz}\gtrapprox0.3$. The occurrence  of perfect absorption in thin, anisotropic ENZ layers is one of the hallmarks
 of coherent perfect absorption\cite{feng}, 
 which couples light to a fast wave propagating along the WS interface. 
A main feature of coherent perfect absorption is
 the  intricate dependence on the real and imaginary parts
 of $\epsilon_{zz}$ which can
lead to the formation of
localized waves inside very narrow regions.
The perfect conductive substrate serves as the reflecting surface that results in
the destructive interference of 
incoming waves,
so that under the proper conditions, 
the  incident beam becomes completely absorbed.

Previously, in Fig.~\ref{ezz_vs_ef}, we found that by tuning the chemical potential, not only can an ENZ response be achieved, but also
that the loss in the WS can be highly sensitive to changes in $\mu$.
To study the effects that variations in $\mu$ have on the EM response of the WS,
in Fig.~\ref{AvEF}, we  
present   the absorptance of the WS structure 
as a function of the normalized $\mu$.
Three different orientations of the incident wave are considered,  as shown in the legend.
We consider a representative value for a  type-II case,  $\beta=1.01$, which according to Fig.~\ref{ezz_vs_ef},
puts the system in the ENZ regime at
$\mu/(v_F|Q|)\approx 0.17$. 
For this value of the chemical potential, the loss corresponds to
$\epsilon''_{zz}\approx 0.3$. 
The remaining components of the permittivity tensor are calculated from 
Eqs.~(\ref{dispexx}) and (\ref{dispgam}), resulting in 
$\epsilon_\parallel \approx 2.8$, and $\gamma \approx 2.9$. 
Note that while conventional absorbers are often 
restricted by their relatively large thicknesses, remarkably
the WS layer 
exhibited here 
has an extremely subwavelength thickness, corresponding to $d/\lambda=1/100$.
Thus, for an incident wavelength of $\lambda \sim 10 \mu{\rm m}$,
this implies $d\sim 100\,{\rm nm}$. We note that the results are 
relatively insensitive to  $\epsilon_\parallel$.
Moreover, for the narrow 
WS widths considered here, the off-diagonal gyrotropic component $\gamma$ 
has a  limited effect on the results,  so that  Eq.~(\ref{r1_gam0}) is often suitable 
for describing the reflection characteristics
over a broad range of parameters.
Overall, we find that
the phenomenon  presented   here are dictated 
mainly by $\epsilon_{zz}$.
This is consistent with the results of Fig.~\ref{AvEF}, where
weak absorption 
occurs for smaller $\theta$, but when $\theta=80^\circ$,
there is complete absorption.
As Fig.~\ref{2dtheta} showed, 
increases in
the dissipative component $\epsilon''_{zz}$ require 
that the incident waves approach the interface at larger $\theta$ 
in order to be
absorbed perfectly.

We now proceed to
show that for certain $\mu$ and tilting $\beta$,
both
type-I and type-II WS systems can completely absorb  the incident 
EM radiation over a relatively wide range of incident wave orientations $\theta$.
In Fig.~\ref{Avtheta},
the absorptance is shown as a function of $\theta$
for a few normalized $\mu$ (see legend).
For chemical potentials
outside of this range,  
the imaginary component $\epsilon''_{zz}$
is either too small or too large to achieve perfect absorption (see Fig.~\ref{ezz_vs_ef}).
The subwavelength slab width is again fixed at 
$d/\lambda=1/100$, and the incident wave 
has a frequency corresponding to
$\omega/(v_F |Q|)=0.3$.
Beginning with the largest chemical potential, $\mu/(v_F|Q|)=0.29$, 
we find that
 perfect absorption occurs
at close to normal incidence. 
This is because as
Fig.~\ref{ezz_vs_ef}(b) showed, 
when $\mu/(v_F|Q|)=0.29$,
a very small amount of loss
is present. 
Therefore,  from  Fig.~\ref{2dtheta}, $\theta$ must be small
in order for the incident beam to couple to the
 EM modes responsible for perfect absorption.
Besides having loss, it is also necessary  for $\epsilon'_{zz}\approx 0$, which  
as in Fig.~\ref{ezz_vs_ef}(a)
 shows, only small  $\beta<1$ in this case
 results in an ENZ response.
 By decreasing the chemical potential, the WS becomes more dissipative.
 Thus we find that each of the perfect absorption peaks in 
 Fig.~\ref{Avtheta} gets shifted towards grazing incidence ($\theta\rightarrow90^\circ$).
 This however requires greater tilting of the Weyl cones to achieve $\epsilon'_{zz}=0$,
 which in some instances, corresponds to a type-II situation where
 $\beta$ exceeds unity (see Fig.~\ref{ezz_vs_ef}).

Finally, to show the importance of using subwavelength WS structures
in the ENZ regime
to achieve perfect absorption,
we  investigate  how
changes in the width $d$ of the WS
(see Fig.~\ref{diagramxz}) affects the absorption properties of the system.
In Fig.~\ref{Avtheta2}, the
color maps depict the absorptance as a  function 
of the normalized width $d/\lambda$ and  incident angle $\theta$.
Both types of WS are considered: (a) type-II with $\beta=1.01$, and (b)  type-I with $\beta=0.96$.
In panel (a) 
$\mu/(v_F |Q|)=0.15$, which corresponds to $\epsilon_{zz}\approx 0.12 + 0.37 i$,
and a gyrotropic parameter of $\gamma\approx3$.
In (b) the chemical potential is increased  to $\mu/(v_F |Q|)=0.28$,  
so that the WS  still has $\epsilon_{zz}$ in the ENZ regime, but with very little loss, corresponding to 
a small imaginary component
$\epsilon''_{zz} \approx 0.005$.
The gyrotropic parameter is relatively unchanged  
 from the previous case, with now $\gamma\approx 3.1$.
For both panels, the normalized frequency is set at  $\omega/(v_F |Q|)=0.3$.
It is evident  that the type-II WS in (a) admits perfect absorption over larger angles,
and that the normalized  width should satisfy  $d/\lambda \lessapprox 1/10$  for appreciable absorption.
As the incident beam is directed  more towards grazing  angles ($\theta\rightarrow90^\circ$), it is apparent that $d$
must be continuously reduced in order for the system to remain a perfect absorber.
For  widths that are larger than the range shown here, 
coupling between the incident beam and the WS system becomes substantially diminished 
as additional reflections are introduced that destroy the previous coherent effects.
For the type-I case (b),  the widths again need to be subwavelength, satisfying $d/\lambda \lessapprox 1/10$,
and
as mentioned above,  perfect absorption arises at
small inclinations of the incident beam due to the weakly dissipative nature 
of $\epsilon_{zz}$ for these system parameters. 
We also see a trend similar to the type-II case in (a), where increases in $\theta$  require thinner WS widths
to achieve $A\approx 1$.

\section{Conclusions}\label{conclusion}

In this paper, we 
 studied the dielectric response of anisotropic type-I and type-II tilted Weyl semimetals. 
 We presented both analytic and numerical results
 that characterized 
 each component of the permittivity tensor. 
 We showed that depending on the  Weyl cone tilt, chemical potential 
 and electromagnetic  wave frequency, the
 component of the permittivity tensor normal to the interfaces can achieve  an epsilon-near-zero (ENZ) response. 
At the charge neutrality point, we showed that 
only type-II Weyl semimetals 
can exhibit an ENZ response.
We also discussed how losses near the ENZ frequency can
be controlled and
 effectively eliminated by properly adjusting the Weyl cone tilt and chemical potential. 
 Making use of the calculated 
  permittivity tensor for the Weyl semimetal, 
  we also investigated 
   the electromagnetic response of a Weyl semimetal structure consisting of a
   planar Weyl semimetal adjacent to a perfect conductor  in vacuum. 
   Our findings showed that thin Weyl semimetals with an ENZ response
   can be employed as coherent 
   perfect absorbers for nearly any incident angle, by choosing the proper geometrical and material parameters.

\acknowledgements 
K.H. is supported in part by ONR and a grant of HPC resources from the DOD HPCMP.  
M.A. is supported by Iran's National Elites Foundation (INEF).
A.A.Z. is supported by the Academy of Finland.

\onecolumngrid

\end{document}